\documentclass[%
 aip,
rsi,%
amsmath,amssymb,
preprint,%
reprint,%
]{revtex4-1}

\usepackage{graphicx}
\usepackage{dcolumn}
\usepackage{bm}
\usepackage[colorlinks=true,linkcolor=blue,citecolor=cyan]{hyperref}
\usepackage[version=4]{mhchem}
\usepackage{float}

\usepackage{color}
\usepackage[dvipsnames]{xcolor}
\usepackage{multirow}
\usepackage{rotating}
\usepackage{threeparttable}
\usepackage{braket}
\usepackage{bbold}
\usepackage{inputenc}
\usepackage{caption}
\usepackage{subcaption}

\usepackage{xr}
\usepackage{soul}
\newcommand{\revS}[1]{{\color{black}{#1}}}

\begin{document}

\title{Relativistic EOM-CCSD for core-excited and core-ionized state energies based on the 4-component Dirac-Coulomb(-Gaunt) Hamiltonian}

\author{Lo\"ic Halbert}
\email{loic.halbert@univ-lille.fr}
\affiliation{Universit\'e de Lille, CNRS, UMR 8523 -- PhLAM -- Physique des Lasers, Atomes et Mol\'ecules, F-59000 Lille, France}

\author{Marta L. Vidal}
\email{malop@kemi.dtu.dk}
\affiliation{DTU Chemistry -- Department of Chemistry, Technical University of Denmark, DK-2800 Kongens Lyngby, Denmark}

\author{Avijit Shee}
\email{ashee@umich.edu}
\affiliation{Department of Chemistry, University of Michigan, Ann Arbor, Michigan 48109, USA}

\author{Sonia Coriani}
\email{soco@kemi.dtu.dk}
\affiliation{DTU Chemistry -- Department of Chemistry, Technical University of Denmark, DK-2800 Kongens Lyngby, Denmark}

\author{Andr{\'e} Severo Pereira Gomes}
\email{andre.gomes@univ-lille.fr}
\affiliation{Universit\'e de Lille, CNRS, UMR 8523 -- PhLAM -- Physique des Lasers, Atomes et Mol\'ecules, F-59000 Lille, France}

\begin{abstract}
We report an implementation of the core-valence separation approach to the 4-component relativistic Hamiltonian based equation-of-motion coupled-cluster with singles and doubles theory (CVS-EOM-CCSD), for the calculation of relativistic core-ionization potentials and core-excitation energies.  With this implementation, which is capable of exploiting double group symmetry, we investigate the effects of the different CVS-EOM-CCSD variants, and the use of different Hamiltonians based on the exact 2-component (X2C) framework, on the energies of different core ionized and excited states in halogen (CH$_3$I, HX and X$^-$, X = Cl-At) and xenon containing (Xe, XeF$_2$) species. Our results show that the X2C molecular mean-field approach [Sikkema \emph{et al.} J.\ Chem.\ Phys.\ 2009, \textbf{131}, 124116], based on 4-component Dirac-Coulomb mean-field calculations ($^2$DC$^M$) is capable of providing core excitations and ionization energies that are nearly indistinguishable from the reference 4-component energies for up to and including fifth-row elements. We observe that two-electron integrals over the small-component basis sets yield non-negligible contributions to core binding energies for the K and L edges for atoms such as iodine or astatine, and that the approach based on Dirac-Coulomb-Gaunt mean-field calculations ($^2$DCG$^M$) are significantly more accurate than X2C calculations for which screened two-electron spin-orbit interactions are included via atomic mean-field integrals.
\end{abstract}
\maketitle


\section{Introduction}
\label{Intro}

X-ray spectroscopies, which typically probe core electrons through electronic excitation or ionization, are particularly suitable techniques to study the local environment of atoms, molecules and materials, as the localized nature of the core orbitals makes them very selective and sensitive.~\cite{book:1992_stohr_nexafs,2018_norman_chemrev}

Over the last few years, new X-ray free electron lasers (XFEL) and last generation synchrotrons have started operating. These advanced light sources have opened the door to a variety of new X-ray-based spectroscopies,~\cite{book:2016_lamberti_XRay:,book:2014_mobilio_XRay,2017_bergmann_rsc_XFEL:} including those operating in time-resolved and non-linear regimes.\cite{book:2016_lamberti_XRay:,book:2014_mobilio_XRay,2017_bergmann_rsc_XFEL:,2017_nisoli_chemrev_atto,2018_norman_chemrev}
As stated for instance by Milne, Penfold and Chergui in their review:~\cite{2014_milne_coordchemrev} ``[...] {\em The progress of experimental techniques for core level spectroscopies is unraveling subtle spectral features
implying that high level theoretical approaches are required to interpret them} [...]''.
The experimental X-ray spectra need to be compared to highly accurate theoretical calculations to assign spectral features and to relate experimental measurements to the structure and dynamic properties of the probed molecular system. 

Among the methods one can use to calculate
core binding energies $\Delta$SCF~\cite{PhysRev.139.A619,deltascf:bagus1971} stands out due to the combination of low computational cost and good results. By performing separate SCF calculations on the original (generally neutral, and closed-shell) and ionized (generally open-shell) species, it introduces the important orbital relaxation that accompanies the creation of the core hole, though at the expense of requiring several calculations for the edges of interest, and with the potential complication that the required open-shell calculations may be difficult to converge. The same approach can be applied in a straightforward manner to DFT,~\cite{dft-core:besley2021} with the additional complication that it will show a certain dependence on the chosen density functional. The same idea of performing separate calculations 
for the initial and the target states 
can be used to devise approaches based on  correlated wavefunctions such as MP2~\cite{south2016} or coupled-cluster (CC).~\cite{zheng2019,DeltaCC_DIP_DEE} 
Analogous methods for
core excitations have also been formulated,\cite{MOM-CORE,DeltaSCF_XAS,SGM:ROKS:Core,PSIXAS,2015_derricotte_pccp_simulation,Verma2016,HUNT1969414,AGREN199475,Ferre2002} 
though this has proven somewhat more cumbersome than for ionizations, since one has to select orbital pairs that should represent the transitions, and converge such excited configurations.
In a number of these approaches the orbitals for the ground and excited/ionized states are not orthogonal, making for instance the calculation of transition moments more involved.

The difficulties above are circumvented with approaches that diagonalize the same Hamiltonian for both the original and the ionized or excited states. 
As such, excited state approaches based on wavefunction theory, like, for instance, (multireference) configuration interaction (MRCI),~\cite{Lundberg2019,mcscf:bokarev2020,KLOOSTER2012122,Maganas2019_MREOMCC} equation of motion CC (EOM-CC)\cite{2012_coriani_pra,2015_coriani_jcp,CVS:erratum,2017_sadybekov_jcp_CoreIE,2019_vidal_jctc,2015_peng_jctc,newCC3,Park:2019,Matthews2020} and multireference (EOM)CC,\cite{Brabec2012,Maganas2019_MREOMCC}
and linear response complete active space SCF (LR-CASSCF),~\cite{xas:helmich-paris2020}
on Green's functions such as the algebraic diagrammatic construction (ADC),~\cite{2001_schirmer_jcpa_ADC_ISR_Kshell,2014_wenzel_jcc,2015_dreuw_wires_ADC_review,2016_neville_rsc_ADC_TRNEXAFS}
or on time-dependent density functional theory (TDDFT),~\cite{dft-core:besley2021,PhysRevLett.97.143001,besley:dft:tddft:core} can all be used to target core excited or ionized states. One downside is, however, that now relaxation has to be accounted for by the correlated electronic structure approach. Among these, single-reference CC methods, and in particular those based on EOM and the closely related linear response (LRCC) formalisms,~\cite{1993_stanton_EOMCC,2012_Sneskov_wires_EOMRev,2012_bartlett_wires_CC_EOMCC,2008_krylov_annrevphyschem_eom,1998_christiansen_ijqc,1990_koch_jcp,EOMRESPONSE} 
are capable of treating electron correlation accurately and in a balanced way between ground and excited states, including a good deal of relaxation effects,
and 
have
an appealing ``black box'' nature (though they are less adept than MRCI/MRCC at treating cases for which the ground state shows a strong multireference nature).

Irrespective of the correlated method, a particular difficulty for determining core states by diagonalizing a many-body Hamiltonian is that such states appear at high energies, where spectra are potentially very dense (see for example figure~\ref{fig:I-_full_Spectrum}). Thus, a naive application of procedures which target exterior roots (such as the Davidson algorithm), would require solving for a very large number of roots. 
One way to overcome this issue is through the use of methods capable of solving for interior roots without requiring that all of them are targeted, as those recently devised and applied to multiconfigurational,~\cite{xas:helmich-paris2020}
EOMCC~\cite{2015_zuev_jcc_solvers,2015_peng_jctc} and mean-field~\cite{iVI:huang2017,iVI:huang2019} wavefunctions.

A more widespread strategy,~\cite{STENER2003115,2001_schirmer_jcpa_ADC_ISR_Kshell,2014_wenzel_jcc,2015_dreuw_wires_ADC_review,2015_coriani_jcp,CVS:erratum,2019_vidal_jctc,Bruno_gs_Xray_ionization,CVS:DFT:MRCI,Lundberg2019,xas:helmich-paris2020,manystates:delcey2019,Matthews2020,Park:2019}
which we follow in this paper, is to employ physically motivated approximations to the original Hamiltonian, in particular within the core valence separation (CVS) approximation.\cite{1980_Cederbaum_PRA_CVS} The core-valence separation is justified by the large spatial and energetic separation between valence and core orbitals/electrons. On the one side, CVS allows to straightforwardly address the highly-energetic core states with minor modifications to preexisting eigenvalue solvers. On the other side,
the CVS helps alleviating convergence issues due to the large number of valence ionized states in the frequency region of the core excitations, which is particularly afflicting methods that explicitly include double excitations in their parametrization.~\cite{Faber:2020,Nanda:2020} 
One can also regard the core states as metastable
Feshbach resonances embedded
in the (valence) ionization continuum, and application of CVS helps stabilizing them.
Further details on the CVS approach will be given in section~\ref{sec:theory:implementation}.

A second issue to be addressed for core states is the importance of relativistic effects,~\cite{book:2007_dyall_faegri,2011_saue_cpc_hamprimer,relativity:liu2020} which significantly alter the energies of inner electrons, and consequently the core spectra of atoms and molecules: due to the high velocities of their electrons (a significant fraction of the speed of light), core $s$ and $p$ orbitals contract (with an associated lowering of their orbital energies) while, due to increased screening, $d$ and $f$ orbitals expand (with an associated increase in their orbital energies). At the same time, orbitals which are degenerate in a non-relativistic framework ($p$, $d$, $f$) are (strongly) split due to spin-orbit coupling, and particularly so for the innermost orbitals. 

Relativistic effects are so pronounced in the core region that, even for molecules containing only elements of the first and second rows, they are very important for the accurate determination of K and L edge spectra. While rather approximate treatments of relativistic effect can yield accurate results for light elements,~\cite{zheng2019,Carbone2019,2019_liu_jctc,vidal2020,keller2020} a more consistent way of  treating these effects is through an electronic Hamiltonian based on the 4-component Dirac operator, as it is accurate all the way down to the heaviest elements (5d and 6d transition metals, lanthanides and actinides). As such, 4-component based methods are ideally suited to treat the core spectra across the periodic table, and to probe K, L and M edges of heavy elements,~\cite{2007_fillaux_jac_actinide,2013_bagus_surfscirep,south2016} which are much more complex to interpret than those of lighter elements. Though the availability of efficient implementations~\cite{Repisky2020,belpassi2020,zhang2020} has in recent years enabled mean-field 4-component calculation for large-scale applications, the additional basis sets required to represent the small component part of the molecular spinors still places a considerable burden on such calculations compared to the non-relativistic case. 

An important development to overcome such issues has been the introduction of the so-called eXact 2-Component (X2C) methods~\cite{x2c:kutzelnigg2005,jensen:rehe2005,x2c:atom2mol:liu2006,Ilias2007,x2c:atom2mol:peng2007,x2c:liu2009,2009_sikkema_jcp}, in which a transformation to decouple the positive and negative energy states of the Dirac Hamiltonian is available in matrix form and, unlike its more approximate counterparts, yields exactly the same  positive energy spectrum as the original 4-component Hamiltonian. We refer the reader to a recent, comprehensive review on relativistic electronic structure for more details~\cite{relativity:liu2020}. 

We note that among the different X2C variants, those that construct the transformation matrix on the basis of converged 4-component atomic or molecular mean-field calculations~\cite{x2c:atom2mol:liu2006,x2c:atom2mol:peng2007,2009_sikkema_jcp} are particularly interesting to correlated calculations~\cite{2009_sikkema_jcp,mfso:cc:liu2018,x2camf:cc:liu2018,mfso:eomcc:cheng2019,x2camf:cc:asthana2019}, since they avoid the handling of two-electron integrals over small component basis functions in the transformation to molecular spinors, while largely avoiding errors with respect to the original 4-component Hamiltonian on the description of two-electron spin-orbit coupling contributions, compared to approaches in which the decoupling is performed based on the one-electron Dirac Hamiltonian prior to the mean-field step~\cite{Ilias2007}.

In this study, we present an approach to study core ionization (and excitation) through the use of projectors, inspired by the work of~\citeauthor{2015_coriani_jcp},\cite{2015_coriani_jcp,CVS:erratum} but extended to the 4-component EOM-CCSD approach developed by~\citeauthor{2018_shee_jcp}~\cite{2018_shee_jcp} in the \textsc{Dirac} relativistic electronic structure package.~\cite{saue2020} Because of the availability of different Hamiltonians in \textsc{Dirac}, we will also investigate the performance of 2-component approaches~\cite{2009_sikkema_jcp} with respect to the 4-component one. We refer to, e.g., Refs.~\citenum{Pathak:2016ck,Pathak:2015he,Pathak:2014gv,Pathak:2014dm,Pathak:2016dk,Klein:2008hx,Yang:2012gd,Wang:2015jw,Epifanovsky:2015hsa,Cao:2016fx,Cao:2017id,Zhang:2017jj,Akinaga:2017hv,Wang:2016hx,minggang2020,himadri2020} for other examples of relativistic EOM-CC implementations, and to Refs.~\citenum{2019_vidal_jctc,Faber:2020,Bruno_gs_Xray_ionization,eTpaper,Nanda:2020,2019_liu_jctc,zheng2019,Park:2019,Sorensen_2020,Matthews2020} for other examples of CVS-EOM-CC implementations.
We also wish to underline that the literature on the theoretical approaches for core spectroscopy is vast and rapidly increasing, and the works cited in the previous paragraphs cannot be considered exhaustive. We refer the reader to a number of recent review papers for more thorough accounts.~\cite{2018_norman_chemrev,besley:dft:tddft:core,mcscf:bokarev2020,Penfold:review:2021}

The paper is organized as follows: In Section \ref{sec:theory}, we give a brief outline of EOM-CCSD (\ref{sec:theory:eomcc}) and of CVS (\ref{sec:theory:CVS}), as well as the details of the current implementation (\ref{sec:theory:implementation}). In Section \ref{sec:computational_details}, we provide the computational details of the calculations. In Section \ref{sec:results}, we present the results obtained with the newly implemented method, where we discuss the accuracy of the CVS approximation (\ref{sec:results:cvs-error}), the performance of the different CVS variants (\ref{sec:results:cvs-variants}), the influence of the Hamiltonian (\ref{sec:results:hamiltonian}) and the comparison to experiment for ionization (\ref{sec:results:ip}) and excitation (\ref{sec:results:ee}) energies. Finally, Section \ref{sec:conclusions} summarizes our conclusions.

\section{Methods and Implementation} \label{sec:theory}

In what follows, indices $(i,j,k,l)$, $(a,b,c,d)$ and $(p,q,r,s)$ refer to occupied, virtual and general orbitals, respectively.

\subsection{EOM-CC} \label{sec:theory:eomcc}
In EOM-CC, the ground-state is treated at the coupled-cluster level. Its eigenfunction is hence given by the exponential ansatz:
\begin{equation}
  |\Psi_\text{CC}\rangle=e^{\hat{T}}|\Phi_0\rangle; \quad\hat{T}=\sum_{\mu}t_{\mu}\hat{\tau}_{\mu}
\end{equation}
where $\Phi_0$ is the reference (typically Hartree-Fock) determinant and the operator $\hat{T}$ is the cluster operator. Truncating $\hat{T}$ to single (S) and double (D) excitations yields the coupled-cluster singles-and-doubles (CCSD) model:
\begin{equation}
\hat{T}=\hat{T}_1+\hat{T}_2;\quad\hat{T}_1=\sum_{ia}t_i^aa_a^{\dagger}a_i;\quad
\hat{T}_2=\frac{1}{4}\sum_{ijab}t_{ij}^{ab}a_a^{\dagger}a_b^{\dagger}a_ja_i,
\end{equation}

The energy and the cluster amplitudes are found from the CC equations:
\begin{align}
\label{eq:cc_eqs}
\langle\Phi_0|\hat{\bar{H}}|\Phi_0\rangle=&\ E\\
\langle\Phi_{\mu}|\hat{\bar{H}}|\Phi_0\rangle=&\ 0;\quad |\Phi_{\mu}\rangle=\hat{\tau}_{\mu}|\Phi_{0}\rangle,
\end{align}
where the similarity-transformed Hamiltonian $\hat{\bar{H}}$ has been defined as:
\begin{equation}
  \hat{\bar{H}}\equiv e^{-\hat{T}} \hat{H}e^{\hat{T}}.
\end{equation}
In the equation-of-motion coupled-cluster (EOM-CC) method\cite{1993_stanton_EOMCC,2008_krylov_annrevphyschem_eom} the target states are obtained
by the diagonalization of the non-Hermitian similarity-transformed Hamiltonian. This non-Hermicity gives rise to right ($R$) and left ($L$) eigenvectors that are not adjoints of each other, obtained solving two different eigenvalue equations:
\begin{equation} \label{Eq:right}
\hat{\bar{H}} |R_{\mu}\rangle = E_{\mu} |R_{\mu} \rangle
\end{equation}
\begin{equation}
\langle L_{\mu}| \hat{\bar{H}} =  E_{\mu} \langle L_{\mu}|, \label{Eq:lambda} 
\end{equation}
for a given excited state ${\mu}$ with energy $E_{\mu}$, where the eigenstates are chosen to satisfy the biorthogonality condition:
\begin{equation}
\langle L_{\mu}|R_{\nu}\rangle = \delta_{\mu\nu}. \label{Eq:biorthogonality} \end{equation}
Right and left  wavefunctions of the target states have been thus obtained from a linear parametrization of the reference state through the $\hat{R}$ or $\hat{L}$ operators
\begin{equation}
|\Psi_{\mu}\rangle = e^{\hat{T}}\hat{R}_{\mu}|\Phi_0 \rangle
\end{equation} 
and
\begin{equation}
  \langle \bar {\Psi}_{\mu} | = \langle \Phi_0 | \hat{L}_{\mu}e^{-\hat{T}}.
\end{equation}
Therefore, the choice of the $\hat{R}$ and $\hat{L}$ operators defines which target states to study, yielding different EOM-CC models. In the present work we focus on the models for electronically excited (EE) states and ionization potentials (IP):
\begin{itemize}
\item EOM-CCSD-EE:
\begin{equation}
\hat{R}^{{\rm EE}} = r_0 + \sum_{ia} r^a_i \{ a^{\dag}_a a_i^{\phantom{\dag}} \} + \sum_{i>j,a>b} r^{ab}_{ij} \{ a^{\dag}_a a^{\dag}_b a_j^{\phantom{\dag}} a_i^{\phantom{\dag}} \} 
\end{equation}
\begin{equation}
\hat{L}^{{\rm EE}} = l_0 + \sum_{ia} l_a^i \{ a^{\dag}_i a_a^{\phantom{\dag}} \} + \sum_{i>j,a>b} l_{ab}^{ij} \{ a^{\dag}_i a^{\dag}_j a_b^{\phantom{\dag}} a_a^{\phantom{\dag}} \}
\end{equation}
\item EOM-CCSD-IP: 
\begin{equation}
\hat{R}^{{\rm IP}} =  \sum_i r_i \{a_i^{\phantom{\dag}}\} + \sum_{i>j,a} r_{ij}^{a} \{a^{\dag}_a a_j^{\phantom{\dag}} a_i^{\phantom{\dag}}\} 
\end{equation}
\begin{equation}
\hat{L}^{{\rm IP}} =  \sum_i l^i \{a^{\dag}_i\} + \sum_{i>j,a} l^{ij}_{a} \{a^{\dag}_j a^{\dag}_i a_a\} 
\end{equation}
\end{itemize}
where curly brackets refer to normal ordering with respect to the Fermi vacuum defined by the reference $\Phi_0$, and the sets $\{\boldsymbol{r}\}, \{\boldsymbol{l}\}$   
to the amplitudes of the corresponding operators.

We have here truncated our $\hat{R}$ and $\hat{L}$ operators at the singles-doubles level since the same truncation is used for the $\hat{T}$ operators.

\subsection{The core-valence separation approximation}\label{sec:theory:CVS}

The essence of the core-valence separation (CVS) approximation\cite{1980_Cederbaum_PRA_CVS} is to decouple valence and core electrons based on their difference in energy and spatial extension. This allows to solve the regular, in this case, EOM-CC equations only in the space of the relevant orbitals.
 However, different flavours of CVS exist, which introduce different levels of approximation. Coriani and Koch first introduced the CVS approximation within coupled-cluster theory\cite{2015_coriani_jcp,CVS:erratum} by applying a projector that zeroes out the amplitudes of all excitations that do not involve at least one core electron. Recently, the frozen-core (fc) CVS-EOM-CCSD approach has been proposed that introduces a further approximation, in this case at the ground-state level: the core orbitals are frozen when solving the CC equations for the ground-state whereas they are the only active ones when solving the EOM equations. Both schemes retain the contribution from excitations simultaneously involving two core orbitals. However, these excitations are located in a much higher range of energy in the spectrum and might therefore also be decoupled. This is the strategy adopted for instance by Dreuw and co-workers in their implementation of the CVS within the ADC family of methods.\cite{2015_dreuw_wires_ADC_review,2014_wenzel_jcc,2015_wenzel_jcp}

In this work, the CVS implementation has been carried out following the recipe proposed by Coriani and Koch:\cite{2015_coriani_jcp} a projector $P$ is applied at each iteration of the Davidson procedure during the resolution of the EOM-CC equations. This projector $P$ selectively zeroes out the unwanted contributions to the EOM trial/solution vectors according to the approximation used. In the CVS approach, unwanted contributions generally correspond to excited determinants involving only occupied valence ($v$) spinors. In the case of EOM-EE, the application of a projector $P\equiv P^\text{CVS}$ to a target electronic state $|\Psi^{{\rm{EE}}}_\mu\rangle$ corresponds to
\begin{equation}
 P^\text{CVS}|\Psi^\text{EE}_\mu \rangle \implies 
 {^\mu}r_i^a = {^\mu}r_{ij}^{ab} = 0 \text{ if } i,j \in v~~,
\end{equation}
though one can also devise variants in which additional contributions are zeroed out, see Section~\ref{sec:theory:implementation} for details. With this definition, the eigenvalue equation to be solved in order to get the energy of the target states becomes:
\begin{equation}
P^\text{CVS} \left(\hat{\bar{H}} P^\text{CVS} |R_{\mu}\rangle\right) = E_{\mu} P^\text{CVS} |R_{\mu} \rangle
\end{equation}

It should be noted that the projection scheme does not yield any savings, in terms of memory usage or operations. 
Some of us\cite{2019_vidal_jctc} have recently gone beyond the use of projection operators and introduced an analytical formulation of CVS, in which only the sub-blocks of the similarity-transformed Hamiltonian that are relevant to the calculation of a given core level are actually formed,  thus resulting in a more efficient implementation. This formulation is naturally combined with the frozen-core approximation separating valence and core spaces. However, one may argue it introduces at the same time an additional error from neglecting core correlation.\cite{2019_vidal_jctc,Sorensen_2020} 

\subsection{Implementation details}\label{sec:theory:implementation}

In this work we focus on exploring the definition of different CVS variants and on assessing their performance following the projection-based scheme, due to its ease of implementation, and with it in view of later implementing the analytical CVS~\cite{2019_vidal_jctc} formulation in \textsc{Dirac}.

We have based our implementation on a flexible scheme to define projection operators. 
\begin{enumerate}
    \item[(a)] A mapping is defined between the excited determinants (and the virtual and occupied spinors' energies associated with each) and the position of each excited determinant in the storage of the ground-state CC amplitudes and EOM-CC coefficients (in the RELCC module of \textsc{Dirac} these are stored in triangular form, and blocked by symmetry, for details see~Refs.\citenum{Visscher1996,Pernpointner2003,Shee_CCgrad2016}); 
    \item[(b)] Information is gathered on how the excited determinant space $\{v\}$ will be treated, i.e.\ whether retained ($P^v=1$) or projected out ($P^v=0$). We have made two options available to the user, based on spinor energies $\epsilon$: ($i$) restricted excitation windows (REWs) for the occupied and virtual, which are defined by setting respective lower ($\epsilon_\text{L}$) and upper ($\epsilon_\text{H}$) bounds for the spinor's energies ;  and ($ii$) CVS, via a single energy that acts as the threshold ($\epsilon_\text{CV}$) for the separation between core and valence;
    \item[(c)] Using the information from the previous steps,  suitable one-particle (singly ionized or excited) unit trial vectors for the core states are then generated, using the values of the diagonal of $\bar{H}$ within the subspaces defined in (b). 
\end{enumerate}
It is important to note that in our implementation, tracking the core excited or ionized states requires the use of a root homing procedure, in which new trial vectors are created by maximizing their overlap with the preceding ones.

In addition to CVS and REW for the excited states, we also used the projection setup to implement the frozen-core approximation, such that we can project out the amplitudes corresponding to core ($c$) orbitals at each iteration during the resolution of the ground-state amplitude equations:
\begin{equation}
 P^\text{core}|\Psi_\text{CC}\rangle \implies 
 t_i^a = t_{ij}^{ab} = 0 \text{ if } i,j \in c
 \label{frozen-core-projector}
\end{equation}
As for the excited states, this projector is defined in terms of spinor energies, via a single threshold or upper and lower bounds defining a window. Finally, it should be noted that the thresholds defining the ground-state and excited state projectors are independent.

\section{Computational Details} \label{sec:computational_details}

All coupled-cluster calculations were carried out with the \textsc{Dirac} electronic structure code~\cite{saue2020} (with the DIRAC19~\cite{DIRAC19} release and revisions \texttt{dbbfa6a, 0757608, 323ab67, 2628039, 1e798e5, b9f45bd}). The Dyall basis sets\cite{2002_dyall_TCA_basis,2003_dyall_TCA_basis,2016_dyall_TCA_basis} of triple-zeta quality (dyall.acv3z) were employed for all species. For selected calculations, on heavy atoms only, we 
performed extrapolations to the complete basis set limit by also considering quadruple-zeta quality Dyall (dyall.acv4z) basis sets. In addition to the Dyall basis sets, the ANO-RCC basis~\cite{ano-rcc-main} were employed for Xe and XeF$_2$. The basis sets were kept uncontracted in all calculations.

Unless otherwise noted, all occupied and virtual spinors were considered in the correlation treatment. 

Apart from the Dirac--Coulomb ($^4$DC) Hamiltonian, for selected calculations we investigated:
\textbf{(a)} the molecular mean-field\cite{2009_sikkema_jcp} approximation to DC ($^2$DC$^M$) and the Dirac--Coulomb--Gaunt ($^2$DCG$^M$) Hamiltonians. For the latter, the Gaunt-type integrals are explicitly taken into account only during the 4-component SCF step, due to the fact that the transformation of these to MO basis is currently not implemented;
\textbf{(b)} the DC Hamiltonian with projecting out all negative-energy solutions of the bare-nucleus one-electron Dirac Hamiltonian from the molecular spinor space, which corresponds to the Furry basis for QED and no-pair Hamiltonians~\cite{PhysRev.81.115} ($^4$DC$^{PF}$); and
\textbf{(c)} the eXact 2-component (X2C)~\cite{Ilias2007} Hamiltonian,
in which we include two-electron spin-orbit contributions via to the untransformed two-electron potential via atomic mean-field contributions calculated with the \textsc{AMFI} code~\cite{HE1996365,marian2001,prog:amfi}, namely spin-same-orbit (X2C-AMFI) and spin-same-orbit and spin-other-orbit (X2C-G-AMFI).

Unless otherwise noted, we employed the usual approximation of the energy contribution from $\left(SS|SS \right )$-type two-electron integrals by a point-charge model.~\cite{1997_visscher_TCA_ssss_int_approx} As it will be discussed below, this approximation introduces negligible errors ($< 0.01$ eV) for lower-energy edges. However, for the higher-energy edges of the heavier systems, 
its error becomes important.

The \ce{XeF2} structure was taken from Ref. \citenum{2013_Cheng_JCP_SR_NMR_SF-2XC}. It corresponds to an optimised structure obtained at   the  SFX2C-1e/coupled-cluster  single  double  triple  (CCSD(T))/unc-atomic  natural orbital (ANO)-RCC level ($r_\text{Xe-F}~=~1.9736$~\AA). The coordinates of \ce{CH3I} come from experimental data in Ref.\citenum{Kuchitsu1998}. Finally, the coordinates for HX were taken from Ref.\citenum{Visscher1996b}.

The dataset can be retrieved at the Zenodo repository.~\cite{halbert2002dataset}

\subsection{Approximations used in CVS}\label{sec:comp-details:cvs-approximations}

Beside the original CC-CVS approach,~\cite{2015_coriani_jcp} we have in addition investigated other approximations:
\begin{enumerate}
    \item Further restricting the definition of the projectors to also drop excited configurations with two core indices (ND, for ``No Doubly core hole determinants''); this corresponds to the method suggested, e.g., in Ref. \citenum{2015_dreuw_wires_ADC_review}, making
    \begin{equation}    
        {^\mu}r^{ab}_{ij} = 0 \text{ if } i,j \in c
   \end{equation}
    \item Freezing the core in the ground-state calculation, retaining thus only the valence orbitals, typically defined as the orbitals with the highest principal quantum number $n$ (FC-V); this essentially corresponds to using eq.~\ref{frozen-core-projector} while setting $\epsilon_{CV}$ to a fairly high value (for example, in the atom of Xe, $\epsilon_\text{4d} < \epsilon_{CV} < \epsilon_\text{5s}$).
    
    \item Freezing all core orbitals except those that are to be targeted in the EOM step (FC-V-except); for example, this means treating a core spinor $k$, that should make up the most important core-excited configurations for state $\mu$, differently from other core spinors in the ground-state calculation. This corresponds to setting projectors for ground and excited states such that
    \begin{align}
       t_i^a  = t_{ij}^{ab} &= 0 \text{ if }\begin{cases}  i,j \neq k \\  i,j \in c \end{cases} \\
       {^\mu}r_i^a = {^\mu}r_{ij}^{ab} &= 0 \text{ if } i,j \in v 
       \label{FC-V-except-projectors}
    \end{align}
    \revS{This approach is equivalent to the one of \citeauthor{Sorensen_2020}\cite{Sorensen_2020}}

    \item The definition of the core/valence spaces \textit{follows} that of the CVS definition/threshold, that is, only the core orbitals with same or lower energy than the ones belonging to the edge under investigation are frozen (FC-f). This is the same approach adopted in Ref. ~\citenum{2019_vidal_jctc}, and it corresponds to setting projectors for ground and excited states such that
    \begin{equation}
       \text{ if } i,j \in v 
       \begin{cases} 
       t_i^a  = t_{ij}^{ab} &= 0 \\
       {^\mu}r_i^a = {^\mu}r_{ij}^{ab} &= 0
       \end{cases}
       \label{FC-f-projectors}
    \end{equation}
    
    \item Further restricting the number of frozen core spinors to only those below the ones of the edge of interest (FC-fpMO {\em{frozen core - follow previous MO}}). For example, in the atom of Xe, if we are interested in the M$_5$ edge (3d$_{5/2}$), this would correspond to a frozen core in the ground-state calculation including  the 1s$_{1/2}$ to 3d$_{3/2}$ spinors, and the 4s$_{1/2}$ and higher spinors make up the valence space in the EOM calculations.

\end{enumerate}

The variants ND and FC-V were also employed with REWs. Since the performance of REW variants was always found to be inferior to their CVS equivalents, they will not be discussed here. The results are however available as supplemental information.

\section{Results}\label{sec:results}

As outlined above, apart from the different Hamiltonians at our disposal, the original CVS formulation itself can be modified through a number of different approximations. In order to draw a clearer picture of their interactions, we proceed in a stepwise fashion: first, using the $^2$DC$^M$ Hamiltonian and focusing on core ionizations, we compare (A) the performance of CVS-EOM-CCSD to the original EOM-CCSD; and (B) the impact of the different approximations that can be introduced in the CVS method itself (see Section~\ref{sec:comp-details:cvs-approximations}). 
Then, having established the relative accuracy of CVS, we investigate (C) the effect of the Hamiltonians on the core ionization energies, and finally (D) proceed to a comparison to experiment of the most accurate setup, focusing mostly on core ionization but also discussing selected core excitations (E).

\begin{figure}[H]
    \centering
\includegraphics[width=0.98\linewidth]{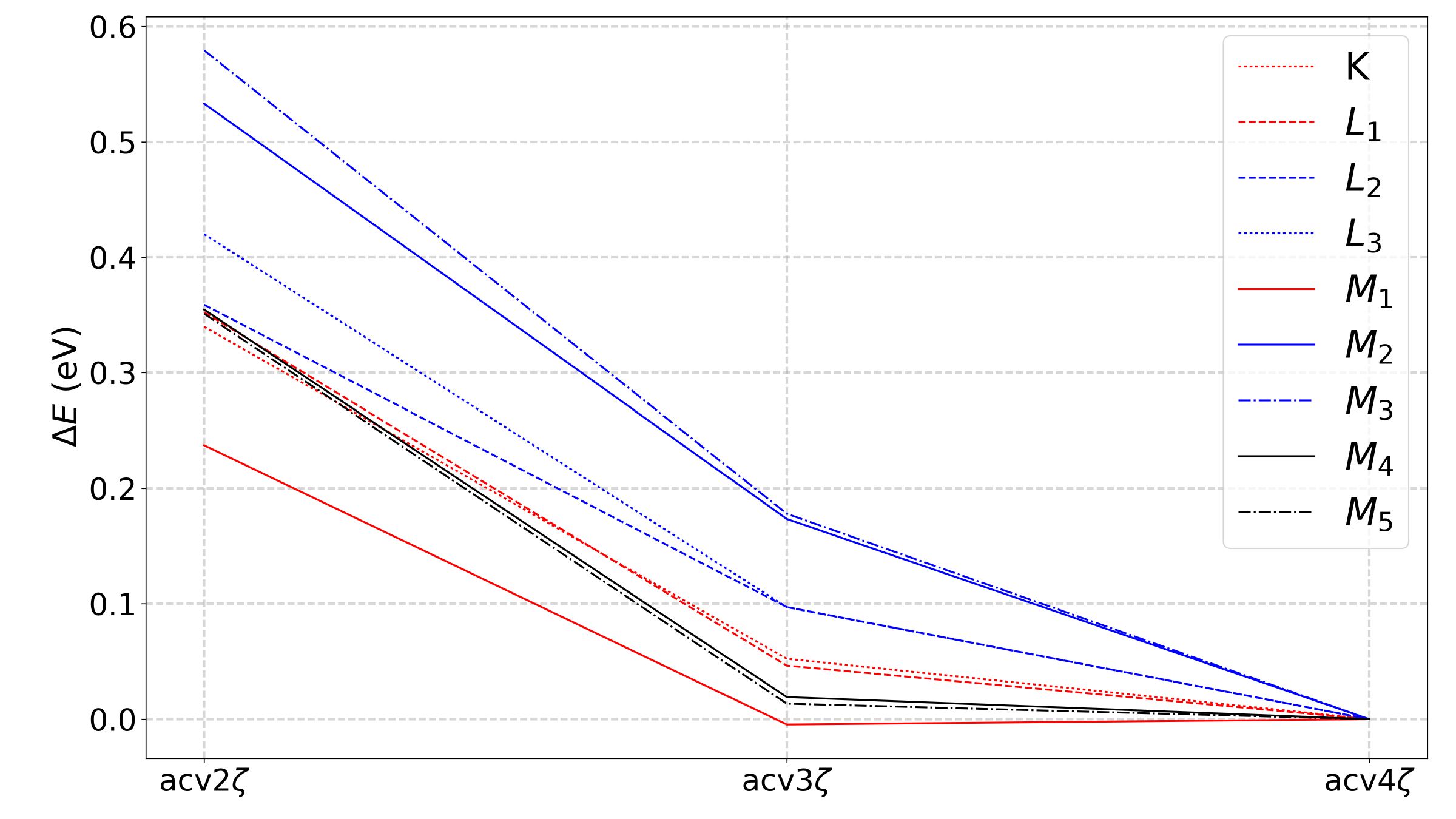}
   \caption{Basis set influence (Dyall basis sets) on the core binding energies of I$^-$ using the CVS approach. All calculations performed with the $^2$DC$^M$ Hamiltonian, and basis sets are kept uncontracted.\label{fig:basis-influence}}
 \end{figure}

In what follows we will not discuss basis set effects. Nonetheless, in Figure~\ref{fig:basis-influence}, we show for I$^-$ a typical behavior: first, there is very little difference between triple and quadruple zeta results (with differences between 0.05 and 0.2 eV), making the corrections due to basis set incompleteness so small that we consider these to be unnecessary. Second, if there are non-negligible differences between the convergence of $s$, $p$ and $d$ shells as the basis sets are improved, the double-zeta basis sets also yield results which are quite good, with differences in binding energies with respect to the triple zeta results no larger than around 0.4 eV, a behavior that may make these smaller basis sets interesting for calculations on larger molecules.

These conclusions are in line with
earlier basis set analyses on light elements.~\cite{Carbone2019,sarangi2020} In the basis set convergence study by~\citeauthor{sarangi2020},~\cite{sarangi2020} in particular, it was shown that, for systems containing only light elements, uncontracted basis sets, even of relatively modest quality for the valence, can provide quite reliable core binding energies. 

\subsection{A comparison of CVS-EOM-IP-CCSD and EOM-IP-CCSD}\label{sec:results:cvs-error}

As mentioned above, the CVS approach provides an efficient and robust way of targeting the low-lying core states. Quantifying the errors introduced by it is therefore an important issue, and we note 
proposals for methods to estimate it. 
\citeauthor{2015_coriani_jcp}\cite{2015_coriani_jcp} evaluated the errors of their CVS variant by comparing to full-space Lanczos results, 
finding deviations of at most a few hundredths of eV.
The authors also proposed a perturbative correction obtained using a L{\"o}wdin partition of the Jacobian and the eigenvalue equation.
A recent study by 
\citeauthor{2020_herbst_jcp_cvs_error}\cite{2020_herbst_jcp_cvs_error}
shows that the CVS error is small and stable across multiple systems within the algebraic diagrammatic method (ADC). 
Thus, we expect to find the same trend for CC. This article also reports the implementation of a post-processing step which removes the error to assess its significance.

Here, we can harness the ability of \textsc{Dirac} to exploit linear symmetry,
and directly compare, for selected species, the CVS-EOM-IP-CCSD states with those obtained by the full diagonalization of the different symmetry blocks of the similarity-transformed Hamiltonian.

\begin{figure*}[ht!]
    \centering
\includegraphics[width=0.98\linewidth]{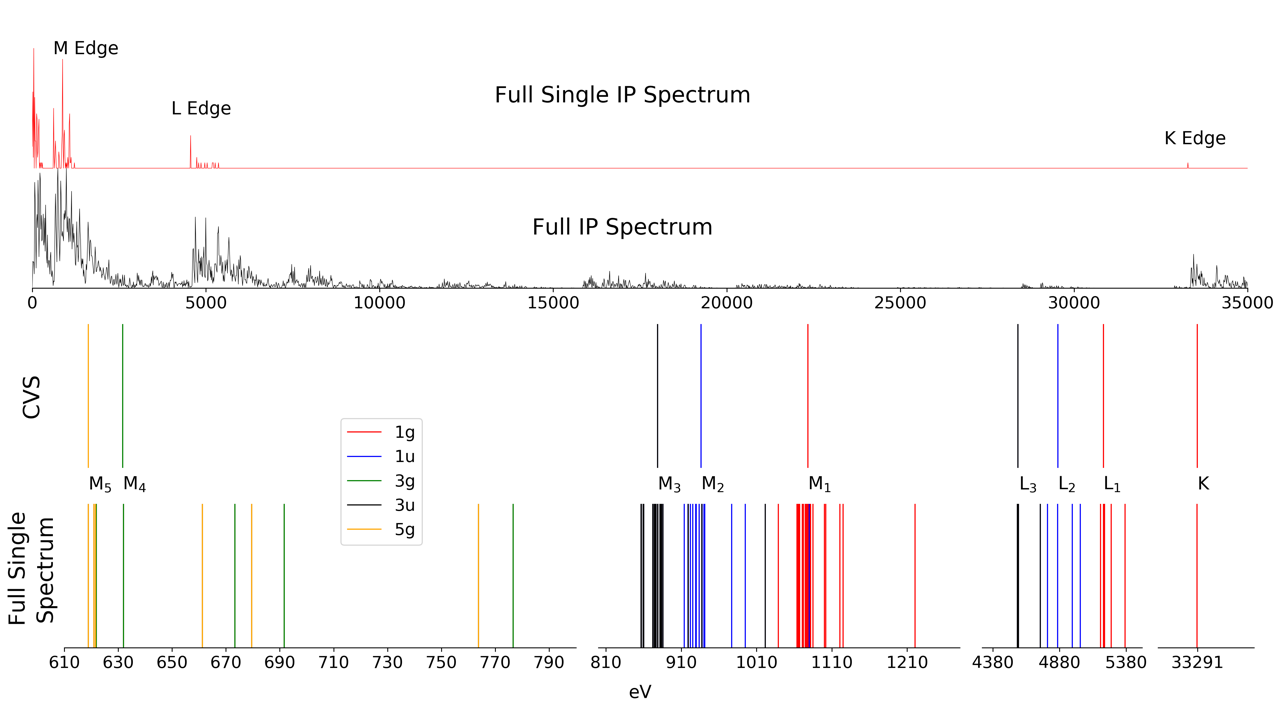}
   \caption{Core ionization energies of iodide (I$^-$) from full diagonalization of $\bar{H}$. Top: Representation of the eigenspectrum up to 35 keV, separated between all states (black) and those with non-negligible singly ionized contributions (red). Values in ordinate correspond to the number of states within a 10 eV interval, with the largest value re-scaled to one. Bottom: Singly ionized states obtained by full diagonalization of $\bar{H}$ and by the CVS approach. The labels $n$g/$n$u indicate the absolute value for the projection of the angular momentum ($|m_j| = n/2, n = 1, 3, 5)$, for the components of spinors of gerade (g) or ungerade (u) symmetry.  All calculations performed with the $^2$DC$^M$ Hamiltonian.
   \label{fig:I-_full_Spectrum}}
 \end{figure*}

The difference between the two approaches is shown graphically for I$^-$ in Figure~\ref{fig:I-_full_Spectrum}. Due to the very dense spectra of $\bar{H}$, 
in the top half of the figure we do not display individual states but rather the number of states per 10 eV intervals. We further discriminate between all states (black) and those containing contributions from singly ionized configurations contributing to at least 1\% of the total wavefunction (red). 

Upon having a closer look at the singly ionized states at the bottom half of the figure, we can more clearly see how the singly ionized states obtained with the CVS approximation (labelled ``CVS'') closely match a subset of those obtained by the full diagonalization (labelled ``full'') of $\bar{H}$.

We note that the additional states presented for the ``full'' calculation  correspond to states that contain significant singly-ionized character but also small but non-zero $2h1p$ contributions. The presence of such states hint at potential pitfalls when targeting only a subset of such highly excited states for some of the L and M edges--not so much in terms of convergence (we have not encountered particular difficulties in converging our calculations) but rather in terms of assignment and comparison to other theoretical or experimental results. 

\subsection{Performance of the CVS-EOM-CCSD variants}\label{sec:results:cvs-variants}

\begin{figure*}[hbt]
    \centering
\includegraphics[width=0.98\linewidth]{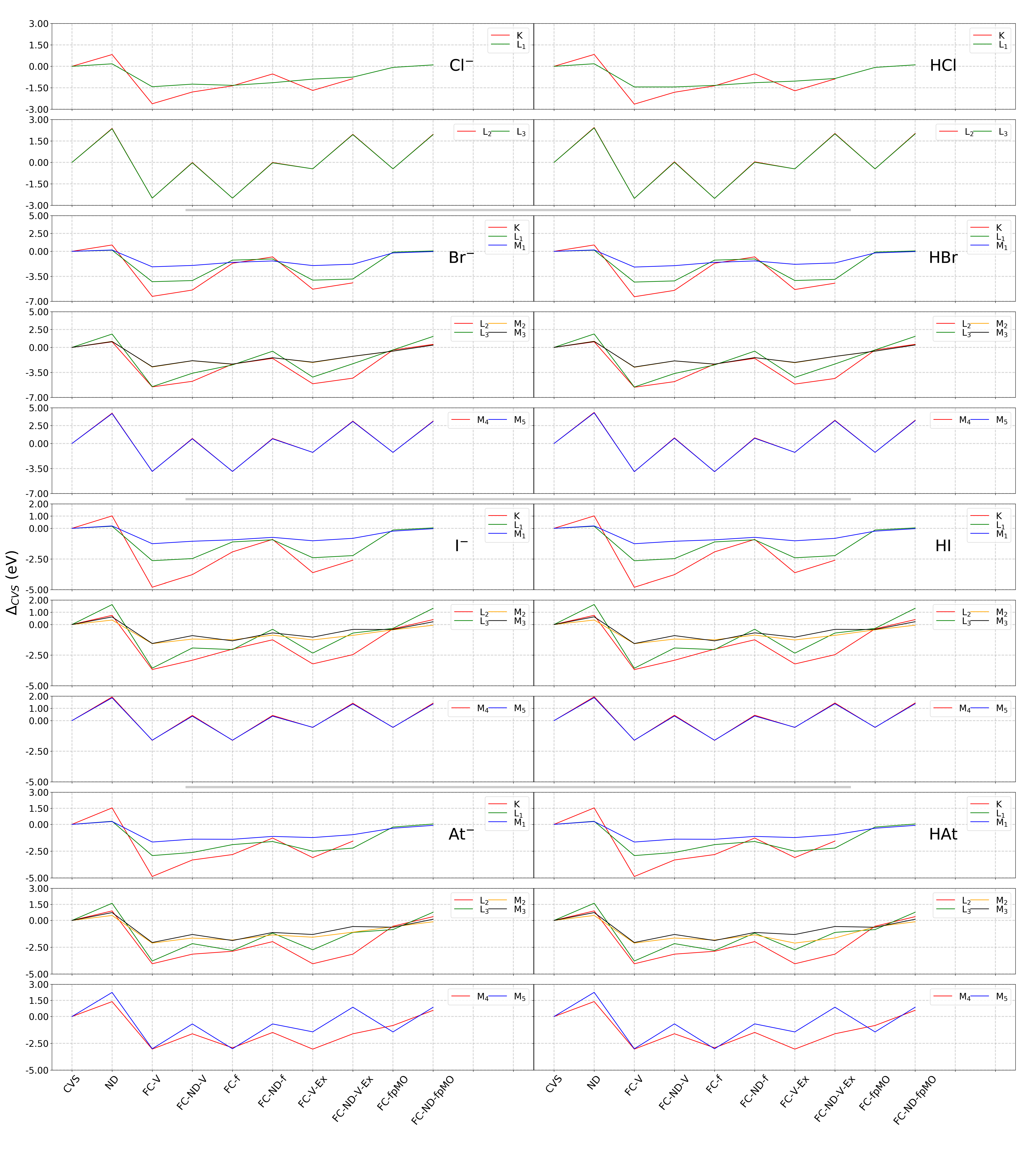}
    \caption{Effect of the CVS variants on the 
    binding energies of X$^-$ and HX systems (X = Cl, Br, I and At). Values (in eV) are relative to the original CVS approach. All calculations performed with the $^2$DC$^M$ Hamiltonian.}
    \label{fig:difference-cvs-variants-anions-hx}
\end{figure*}

Next we turn our attention to the performance of the different approximations that can be employed on top of CVS (see 
Section~\ref{sec:comp-details:cvs-approximations} for their description). Our results for the X$^-$ systems are shown in  Figure~\ref{fig:difference-cvs-variants-anions-hx}. As we see similar trends for Xe, the corresponding figure is shown in the supplemental information (Figure~\ref{SI-fig:Xe}), along with results for the REW approach for the Cl$^-$ to I$^-$
(Figures~\ref{SI-fig:rew:cl},~\ref{SI-fig:rew:br} and \ref{SI-fig:rew:i}) and HCl to HI (Figures~\ref{SI-fig:rew:hcl},~\ref{SI-fig:rew:hbr} and \ref{SI-fig:rew:hi}).

\subsubsection{Freezing core orbitals in the ground-state} 

Freezing the core orbitals in the ground-state CCSD calculation results in a significant lowering of the core ionization energies. This lowering is considerable if the frozen core is taken to represent all subvalence spinors (FC-V), and it remains non-negligible even if the same threshold is used to define both the frozen orbitals and the core-valence separation region (FC-f).
However, if one does not freeze the aimed edge (FC-fpMO), the ionization energy obtained is much closer to the corresponding CVS one. 

Freezing all subvalence spinors but the ones in the edge we are interested in (FC-V-ex), an approximation that
could provide potentially lower computational cost for systems with large number of core electrons such as the elements in the fourth row and beyond, does not 
perform better than (FC-f)---in fact, the opposite appears to be the case for the inner core orbitals.

\subsubsection{Projecting out double-core ionized configurations}

Projecting out doubly-excited core excited determinants (ND) tends to result in an increase of the core ionization energies, which can be significant though always smaller than the underestimation produced by freezing all the core spinors.

As a result, the combination of these two approximations yields an approximate scheme that much better reproduces the CVS energies, though still with non-negligible discrepancies for the heavier elements. Particularly good results are shown by the (FC-ND-f) combination, for which error compensation yields results very similar to the regular CVS approximation. This can be rationalized because core correlation is not taken into account when using (FC-f), 
whereas the inclusion of the double occupied core orbitals in the EOM step does. Therefore, the combination of both approximations is in fact more consistent.

\subsubsection{Efficiency considerations}

Since by using projection operators we do not save in memory or operation counts, the approximations discussed above are not of strong interest by themselves, and will not be used further since, even in the best combination, they will invariably degrade the performance of the original CVS scheme. 

We can nevertheless use our findings to discuss the potential trade-offs between cost and accuracy, as a guide to efficient implementations such as those proposed in Ref.~\citenum{2019_vidal_jctc} for heavy elements. As discussed then, combining the elimination of doubly core excited determinants with the use of a frozen core (FC-ND-f, which is equivalent to the scheme in Ref.~\citenum{2019_vidal_jctc}) seems to yield a computationally efficient approach with a good error cancellation balance, and this remains the case as one goes down the periodic table.

In the case of heavy elements, it may also be interesting to keep a large frozen core (CVS-FC-ND-V, or CVS-FC-V-Ex) since these introduce non-negligible, but seemingly systematic, errors, with the upside that a large core would translate in potentially large computational savings.

This is illustrated in Table~\ref{SI-tab:operation_count}, where we provide,
\revS{for the representative systems
Cl$^-$ and At$^-$},
the operation counts (without reductions due to point group symmetry) for the construction of EOM-EE and EOM-IP $\sigma$ vectors (based on the expressions from~\citeauthor{2018_shee_jcp}~\cite{2018_shee_jcp}, \revS{see expressions in the supplemental information}), in the case of an analytical implementation of the major approximations suggested in this work (FC-f and ND). As seen, the FC-f approximation yields the biggest reductions in operation counts, in particular for EOM-IP. It is also important to note that for a given edge, the heavier the element, the bigger these reductions are.

\subsection{Influence of the Hamiltonian on the core ionizations}\label{sec:results:hamiltonian}

\begin{figure*}[ht!]
    \centering
\includegraphics[width=0.98\linewidth]{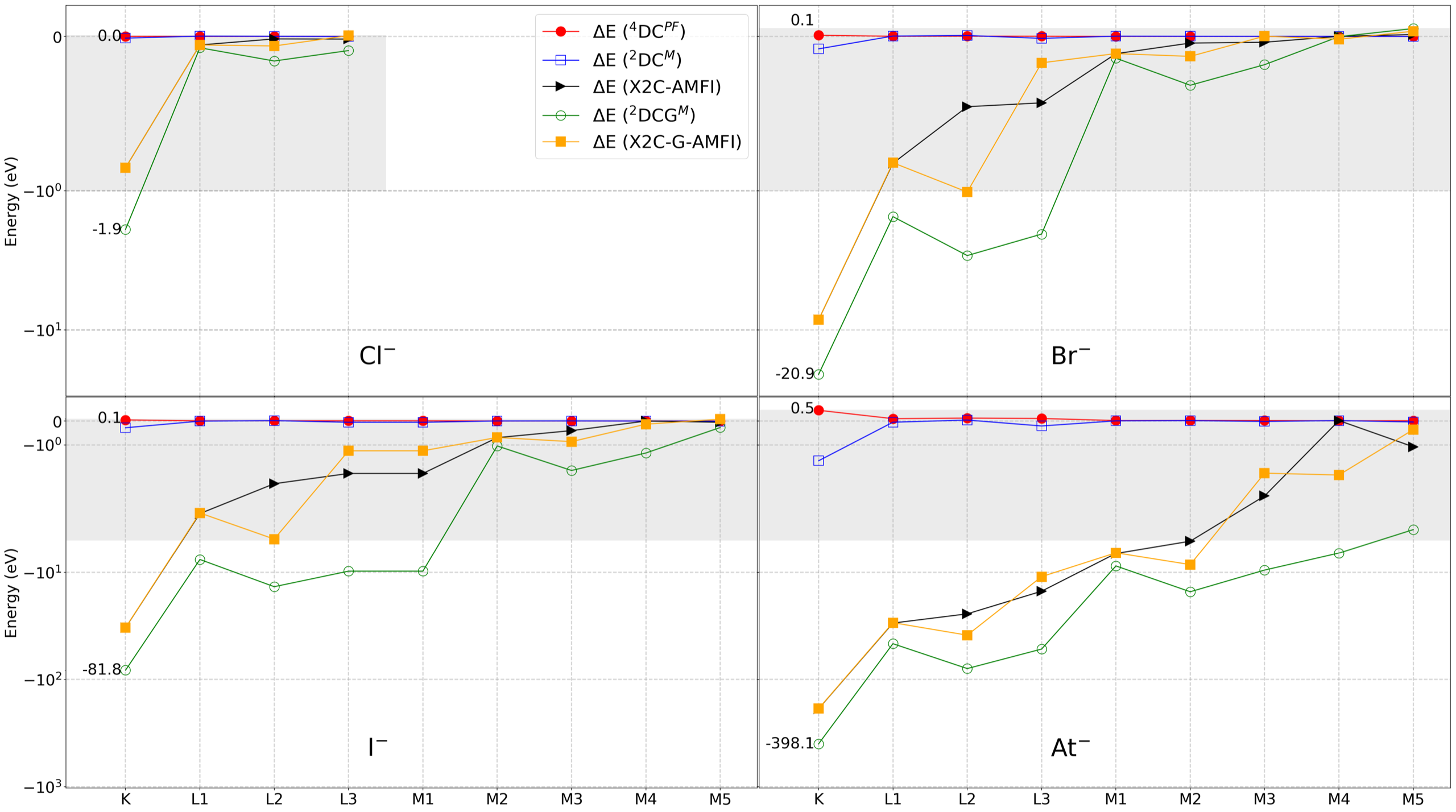}
\caption{Comparison of influence of the Hamiltonian to the core binding energies for different edges, for the halide ions (Cl$^-$ to At$^-$). Instead of binding energies themselves, we present the difference between the binding energies obtained for each Hamiltonian $X$ and {$^4$}\text{DC} ($\Delta E_m(X) = E_m(X) - E_m({^4}\text{DC})$, in eV, $m$ a particular edge). The scales are logarithmic except for the areas in grey for which the scales are linear.}
    \label{fig:difference-hamiltonians}
\end{figure*}

We now turn our attention to the impact of the Hamiltonian on the ionization energies. Our results can be found in Figure~\ref{fig:difference-hamiltonians}, where take the $^4$DC Hamiltonian as reference and we plot the difference in binding energies betweeen it and the other Hamiltonians considered (see figure caption for details), as $^4$DC is the only 4-component Hamiltonian without approximations (apart from the treatment of the $(SS|SS)$ integrals which we address below) we can employ in correlated calculations with \textsc{Dirac}.

First, we observe a very good match between the $^4$DC and $^2$DC$^M$ Hamiltonians across the halogen series---that is, differences in absolute values generally fall below 0.001 eV for the M edges, between around 0.01 and 0.1 eV for the L edges, and are of the order of 0.1 eV for the K edge of all species except astatine, for which the difference is 1.67 eV. This latter discrepancy is still small compared to the K edge binding energy of 96 keV.

To better understand how these discrepancies arise we should recall that in the $^2$DC$^M$ ($^2$DCG$^M$) approximation a calculation with the $^4$DC ($^4$DCG) Hamiltonian is first is carried out for whatever system we are interested in--in general a molecule, hence the denomination ``molecular mean-field'', though in Figure~\ref{fig:difference-hamiltonians} we are in effect dealing with atoms. Upon convergence, the transformation to 2-component is carried out on the Fock matrix itself (so that the 2-component spinor energies correspond exactly to the 4-component ones), but the two-electron operator is left untransformed~\cite{2009_sikkema_jcp,saue2020} and thus introduces a picture-change error in the electron-electron interaction. In valence only calculations, where core-core and core-valence electron correlation is not accounted for, this yields negligible errors for excitation, ionization and electron attachment processes~\cite{2018_shee_jcp}. Our results show that this is still the case for core states of light atoms such as chloride, and for the M and N edges down to the sixth row.

We can numerically assess the effect of using untransformed two-electron operator by comparing how the correlation energy ($E_c$) differ between $^2$DC$^M$ and $^4$DC ($\Delta E_c$), as shown in table~\ref{tab:correlation-energy-comparions}. The $\Delta E_c$ for both MP2 and CCSD yield the same trends, so we shall focus on the simpler MP2 model; as canonical orbitals are employed, and their energies the same for $^2$DC$^M$ and $^4$DC, any differences between the two come from the picture change error in the two-electron integrals. We see that for chloride  $\Delta E_c$ is still relatively small, then it increases five fold going from chloride to bromide, doubles from bromide to iodide, and then increases nearly fourfold from iodide to astatide. 

The effect of the two-electron picture change error can also be seen through a comparison of $^4$DC to a calculation  in which negative-energy solutions of the bare-nucleus one-electron Dirac Hamiltonian are projected out from the molecular spinor space ($^4$DC$^{PF}$), whereby rotations between negative and positive energy states are effectively eliminated. With this, $^4$DC$^{PF}$ spinor energies are slightly different than $^4$DC ones (see dataset~\cite{halbert2002dataset}), but the picture change error in the two-electron integrals is eliminated, and we would thus expect rather similar correlation energies. This picture is consistent with our numerical results, as correlation energies for $^4$DC$^{PF}$ are closer to the $^4$DC ones by at least three orders of magnitude than the $^2$DC$^M$ ones. 

\begin{table}[hbtp]
\centering
\caption{MP2 and CCSD correlation energies as well as differences in correlation energy between $^4$DC ($E_c$ and $\Delta E_c$ respectively, in eV), for different Hamiltonians. X2C(a) refers to X2C-AMFI and X2C(b) to X2C-G-AMFI.~\label{tab:correlation-energy-comparions}}
\begin{tabular}{ll rrrrr}
\hline
\hline
 &     &  $^4$DC$^{PF}$ &  $^2$DC$^M$&  $^2$DCG$^M$ &  X2C(a) &  X2C(b)\\
\hline
Cl$^-$	&	$E_c$(MP2)	&	$-$16.12	&	$-$16.16	&	$-$16.16	&	$-$16.16	&	$-$16.16	\\
	&	$\Delta E_c$	&     $<$2$\times 10^{-6}$	&	$-$0.04	&	$-$0.04	&	$-$0.04	&	$-$0.04	\\
	&	$E_c$(CCSD)	    &	$-$16.43	&	$-$16.47	&	$-$16.47	&	$-$16.47	&	$-$16.47	\\
	&	$\Delta E_c$	&    $<$2$\times 10^{-6}$&	$-$0.04	&	$-$0.04	&	$-$0.04	&	$-$0.04	\\
	&&&&&&\\
Br$^-$	&	$E_c$(MP2)	&	$-$41.44	&	$-$41.65	&	$-$41.64	&	$-$41.65	&	$-$41.65	\\
	&	$\Delta E_c$	&	 $<$5$\times 10^{-5}$	&	$-$0.21	&	$-$0.20	&	$-$0.21	&	$-$0.21	\\
	&	$E_c$(CCSD) &	$-$39.65	&	$-$39.86	&	$-$39.85	&	$-$39.86	&	$-$39.86	\\
	&	$\Delta E_c$	&    $<$5$\times 10^{-5}$	&	$-$0.21	&	$-$0.20	&	$-$0.21	&	$-$0.21	\\
	&&&&&&\\
I$^-$	&	
$E_c$(MP2)	
&	$-$48.54	
&	$-$49.07	
&	$-$49.05	
&	$-$49.06	
&	$-$49.06	\\
	&	$\Delta E_c$	&	  $<$3$\times 10^{-4}$	&	$-$0.53	
	&	$-$0.51	
	&	$-$0.52	
	&	$-$0.52	\\
	&	$E_c$(CCSD)	    &	$-$46.13	&	$-$46.65	&	$-$46.64	&	$-$46.65	&	$-$46.65	\\
	&	$\Delta E_c$	&	  $<$3$\times 10^{-4}$	&	$-$0.52	
	&	$-$0.51	
	&	$-$0.52	
	&	$-$0.52	\\
	&&&&&&\\
At$^-$	&	$E_c$(MP2)	&	$-$90.95	&	$-$92.90	&	$-$92.83	&	$-$92.86	&	$-$92.87	\\
	&	$\Delta E_c$	&	  $<$3$\times 10^{-3}$	&	$-$1.95	&	$-$1.88	&	$-$1.91	&	$-$1.92	\\
	&	$E_c$(CCSD) 	&	$-$84.30	&	$-$86.24	&	$-$86.17	&	$-$86.20	&	$-$86.20	\\
	&	$\Delta E_c$	&	  $<$3$\times 10^{-3}$	&	$-$1.94	&	$-$1.87	&	$-$1.90	&	$-$1.91	\\
\hline
\hline
\end{tabular}
\end{table}

As it will be shown in the following, the sufficiently high accuracy of $^2$DC$^M$ clearly makes it an asset for applications, due to its reduced computational cost in the index transformation step. However, our results indicate two-electron picture change errors are significant enough for deeper cores to require further attention for elements in the sixth row and beyond.

While we do not have $^4$DCG reference values to which compare the approximate Hamiltonians that include the Gaunt interaction such as $^2$DCG$^M$, we expect two-electron picture change errors to generally follow the same trends discussed above. We nevertheless compare $^2$DCG$^M$ to $^4$DC in Figure~\ref{fig:difference-hamiltonians} and Table~\ref{tab:correlation-energy-comparions}, as a way to underscore the importance of the Gaunt interaction for the inner edges. We clearly see that already starting with chloride, we have a non-negligible effect arising from the Gaunt interaction in $^2$DCG$^M$ that lowers the core binding energy, and can amount to nearly 2 eV for the K edge with respect to the $^2$DC$^M$ or $^4$DC. 

We can also compare $^2$DCG$^M$ to the approaches in which the transformation to the 2-component picture is done before the SCF step (X2C-AMFI and X2C-G-AMFI), as opposed to using the full atomic or molecular potential obtained the SCF step as in the $^2$DCG$^M$ approach. We see that the latter qualitatively follow the the changes in binding energy seen for $^2$DCG$^M$ as we move across the rows, though with significant numerical differences for the K edge (with differences of over 40 eV for Iodine and Xenon, and over 200 eV for Astatine) as well as for the L and M edges.

\begin{table}[hbtp]
\centering
\caption{Contributions (in eV) to the core binding energies of I$^-$, Xe and At$^-$ from: (a) the Gaunt interaction and (b) the ($SS|SS$) integral. These contributions are calculated as energy difference between CVS-EOM-CCSD calculations employing (a) the $^2$DCG$^M$ and $^2$DC$^M$ Hamiltonians; and (b) the $^2$DCG$^M$ without and with  inclusion of the ($SS|SS$) integrals at the SCF step.
~\label{tab:gaunt_ssss_contributions}}
\begin{tabular}{l rr p{0.1cm} rr p{0.1cm} rr}
\hline
    & 
    \multicolumn{2}{c}{I$^-$} && 
    \multicolumn{2}{c}{Xe}  &&
    \multicolumn{2}{c}{At$^-$}  \\
\cline{2-3}
\cline{5-6}
\cline{8-9}
Edge         & 
$\Delta E$(a)  & $\Delta E$(b)  &&  
$\Delta E$(a)  & $\Delta E$(b)  && 
$\Delta E$(a)  & $\Delta E$(b)  \\
\hline
K	&	$-$81.57	&	$-$2.58	&&	$-$86.75	&	$-$2.87	&&	$-$396.04	&	$-$40.50	\\
L$_1$	&	$-$7.59	&	$-$0.39	&&	$-$8.15	&	$-$0.44	&&	$-$46.24	&	$-$7.28	\\
L$_2$	&	$-$13.64	&	$-$0.58 &&	$-$14.62	&	$-$0.65	&&	$-$78.99	&	$-$10.46	\\
L$_3$	&	$-$9.70	&	$-$0.21	&&	$-$10.39	&	$-$0.23	&&	$-$51.77	&	$-$3.56	\\
M$_1$	&	$-$1.05	&	$-$0.08	&&	$-$1.14	&	$-$0.09	&&	$-$8.72	&	$-$1.79	\\
M$_2$	&	$-$2.08	&	$-$0.21	&&	$-$2.25	&	$-$0.12	&&	$-$15.19	&	$-$2.35	\\
M$_3$	&	$-$1.34	&	$-$0.08	&&	$-$1.46	&	$-$0.04	&&	$-$9.51	&	$-$0.89	\\
M$_4$	&	$-$0.59	&	$-$0.06	&&	$-$0.66	&	$-$0.03	&&	$-$6.62	&	$-$0.82	\\
M$_5$	&	$-$0.26	&	$-$0.04	&&	$-$0.31	&	
0.01	&&	$-$4.52	&	$-$0.09	\\
N$_1$	&		&		&&		&		&&	$-$1.81	&	$-$0.44	\\
N$_2$	&		&		&&		&		&&	$-$3.29	&	$-$0.55	\\
N$_3$	&		&		&&		&		&&	$-$1.81	&	$-$0.19	\\
N$_4$	&		&		&&		&		&&	$-$0.91	&	$-$0.10	\\
N$_5$	&		&		&&		&		&&	$-$0.42	&	0.02	\\
N$_6$	&		&		&&		&		&&	6.74	&	0.06	\\
N$_7$	&		&		&&		&		&&	0.48	&	0.12	\\\hline
\end{tabular}
\end{table}

Finally, we assess the effect of including the ($SS|SS$) integrals in $^2$DC$^M$ and $^2$DCG$^M$ calculations. Since these contributions are generally very small for elements before the fourth row, and they increase significantly the cost of the SCF step, we have only investigated the fourth and fifth row atoms (I, Xe and At). 

For I$^-$ and Xe, these contributions are relatively modest, with reductions of around 2.6 eV and 3 eV to their K edge binding energies, with other significant reductions for the L (0.4-0.7 eV) and M$_{1,2}$ edges (0.1 eV). We note that there is little variation from these atomic values for the molecular systems (see dataset~\cite{halbert2002dataset}), in line with the much more localized nature of the small component density compared to the large component one. Furthermore, we notice a very subtle difference (0.02 eV) between the $^2$DC$^M$ and $^2$DCG$^M$ Hamiltonians with ($SS|SS$) integrals. 

For At$^-$ the reductions in binding energies are much more significant: around 40 eV for the K edge,  10 eV for the L$_2$ edge, and well above 1 eV for the L$_3$ to M$_2$ edges. Interestingly, these contributions remain around 0.5 eV for the N edges relating to ionizations from $s$ and $p$ spinors, which are comparable to the energies of the M$_1$ and M$_2$ edges for I$^-$ and Xe, which fall between 0.7 and 1 keV.

These results underline the need for explicitly accounting for these integrals (or correcting the energies for their contribution), as soon as we are interested in edges arising from ionizations of $s$ and $p$ spinors for fifth-row elements and beyond.

\subsection{Binding energies: comparison to experiment and prior theoretical works}\label{sec:results:ip}

We now compare our results for the $^2$DC$^M$ and $^2$DCG$^M$ Hamiltonians, including the ($SS|SS$) integrals, to experiment and other theoretical results for the K and L edges of Xe, XeF$_2$ and CH$_3$I, for which recent high energy, gas-phase XPS experiments have been performed (for Xe and XeF$_2$, results are also available for less energetic edges). Our results are found in Table~\ref{tab:comparison-exp-ip}. 

As all our calculations lack contributions from QED effects, and that the $^2$DCG$^M$ does not include the gauge term that is necessary to recover the full (zero-frequency) Breit interaction, we correct our energies with the results from
~\citeauthor{Kozio2018},\cite{Kozio2018}
who provide the values for the Breit and leading QED contributions (self interaction and vacuum polarization) to the atomic spinors of selected closed-shell atoms (among which Xe). The same procedure was followed by~\citeauthor{2019_southworth_pra_kedge}\cite{2019_southworth_pra_kedge}
to correct their calculations, which are based on a combination of 1-component CVS-EOM-CCSD (and CCSDT) calculations with energy estimates for the different  effects (1- and 2-electron scalar relativity, spin-orbit coupling, nuclear size effects).

Taking the Xe atom, the QED effects are most important for the K and L$_1$ edges, lowering the associated binding energies by 42.7 and 5.5 eV, respectively. They are also non-negligible for the L$_3$ and M$_1$ edges, which are lowered respectively by  0.5 and 1.1 eV. The Breit interaction also lowers the binding energies and, apart from the L$_1$ edge (for which both are of the same magnitude) is in general much larger than the QED effects (80.7, 13.1 and 8.8 eV for the K, L$_2$ and L$_3$ edges, respectively) and still important for the M$_3$ edge (1.2 eV).

Since \citeauthor{Kozio2018} only provide the full Breit term, we have estimated the magnitude of the gauge correction from the difference between our $^2$DCG$^M$ results and their Breit values. From that, we have that the gauge term increases the binding energies (6.04 eV for the K edge, 0.51 eV for the L$_1$ edge, and around 1.57 and 1.58 eV for the  L$_2$ and L$_3$ edges, respectively).

We observe our corrected 2-component results differ from experiment by around 2.1 eV, and if we employ the estimate for higher-order correlation contributions to EOM-CCSD from \citeauthor{2019_southworth_pra_kedge},\cite{2019_southworth_pra_kedge} which decreases the binding energies by 3.8 eV, the difference to experiment is now of $-$1.71 eV.  

We have also investigated the use of uncontracted ANO-RCC basis, used by~\citeauthor{2015_southworth_jcp_xef2}~\cite{2015_southworth_jcp_xef2}, which is slightly smaller for $s$ and $p$ primitive sets than the Dyall sets. We obtain differences with respect to experiment of -1.69 eV, which is consistent with our results using the Dyall basis sets. We attribute most of the small differences between ours and prior results to the shortcomings in the treatment of the two-electron interactions in $^2$DC$^M$ or $^2$DCG$^M$ discussed previously, as calculations with $^4$DC yield results in very good agreement with those of~\citeauthor{2015_southworth_jcp_xef2}.  

Beyond the K edge, our corrected calculations compare rather well to the recent experimental results of \citeauthor{Oura2019}~\cite{Oura2019} (5452.7 eV, 5106.7 eV and 4786.7 eV; measurements carried out at BL29XU of SPring-8 in May 2016) with the exception of the L$_1$ edge, for which a larger discrepancy to experiment (7.79 eV) is observed. For the M edges, \citeauthor{2015_southworth_jcp_xef2}~\cite{2015_southworth_jcp_xef2} present experimental results for the M$_{4,5}$ edges, and our corrected calculations differ from experiment by $-$0.94 eV and $-$0.85 eV respectively. For the M$_4$ edge these are quite comparable to the theoretical calculations of Southworth and coworkers,~\cite{2015_southworth_jcp_xef2} now also corrected for QED effects ($-$0.92 eV to experiment), while for the M$_5$ edge both theoretical results differ by around 0.24~eV. 

Given the values for the higher-order correlation effects for the K and M$_{4,5}$ edges, and from the breakdown of relativistic, correlation and QED effects from atomic many-body calculations on Xe at the K, L and M edges~\cite{mooney1992}--which indicate non-negligible differential correlation, relaxation and other effects (for example Auger shifts) for the different L edges, and to a lesser extent for the M edges--we consider future attempts to investigate higher-order electron correlation corrections for the L edges to be of significant interest.

For XeF$_2$, applying the same corrections as above to our $^2$DC$^M$ and $^2$DCG$^M$ calculations in the Dyall basis sets, we arrive at K and M$_{4,5}$ edges edge binding energies differing by roughly $-$0.4~eV from experiment. These differences are smaller but consistent with those obtained for Xe, underscoring the largely atomic nature of these deep core energies. As was the case for the atom, a comparison to $^4$DC results for the K edge indicate that part of the small differences between the results of~\citeauthor{2015_southworth_jcp_xef2} and our 2-component ones come from the shortcomings in the treatment of two-electron interactions.

For CH$_3$I, we have only performed $^2$DCG$^M$ calculations, first because, as illustrated above, there are no significant differences between $^2$DCG$^M$ and $^2$DC$^M$ based results once we account for QED and Breit or gauge contributions. Second, in the $C_s$ point group used, the coupled-cluster wavefunctions are complex-valued, making calculations computationally more expensive. We obtain 33213.79~eV and 5208.55 eV for the K and L$_1$ edge binding energies, values that overestimate the experimental values by 38.59 eV and 11.08 eV. For the K edge, this difference is rather close to the one found for the Xe and XeF$_2$ species. 

As QED and Breit corrections are not provided by~\citeauthor{Kozio2018}~\cite{Kozio2018} for iodine, we have used instead those of~\citeauthor{Boudjemia2019},\cite{Boudjemia2019} which amount, for the K edge, to $-$77.0 eV for Breit, +6.10 eV for the gauge term and $-$39.0 eV for QED, and to $-$7.20 eV, +0.50 eV and $-$5.10 eV for the L$_1$ edge. With these corrections, we now overestimate the K and L$_1$ binding energies by 5.69~eV and 1.95 eV. The missing effect would be that of higher-order correlation corrections. While we cannot estimate this here, we speculate that if it follows roughly what is found for the Xe species, that would decrease the binding energies, and likely take the K edge to a few eV. 

\subsection{Excitation energies: comparisons to experiment and prior theoretical works}\label{sec:results:ee}
As anticipated, the current implementation allows  calculation not only of ionization energies, but also of excitation energies, which will be briefly discussed hereafter.
In this case, since ($SS|SS$) integrals do not significantly affect the energies, only results for the $^2$DCG$^M$ and $^2$DC$^M$ Hamiltonian are presented, without explicit inclusion of the aforementioned integrals.

\begin{table}[H]
\centering
\caption{Comparison between calculated and experimental gas-phase binding energies (in eV) for Xe, XeF$_2$ and CH$_3$I. Calculations are broken down into values obtained with (a) Koopmans theorem, (b) CVS-EOM-CCSD method; (c)  CVS-EOM-CCSD with atomic corrections for QED and Breit (in the case of $^2$DCG$^M$ results, corrections for the gauge term) interactions; and (d) higher-order correlation corrections by Southworth and coworkers on (c). All of our results include contributions from the ($SS|SS$) integrals. Basis sets A: ANO-RCC, D: Dyall\label{tab:comparison-exp-ip}}
\begin{tabular}{ll p{0.1cm} rrrr}
\hline
Edge
& model &
& $E$(a)
& $E$(b)
& $E$(c) 
& $E$(d) \\
\hline
%
\multicolumn{7}{c}{Xe} \\
\hline 
K
& D/$^4$DC &
& 34755.91
& 34690.82
& 34567.44
& 34563.64 \\

& D/$^2$DC$^M$ &
& 34755.91
& 34690.60
& 34567.22
& 34563.42 \\

& D/$^2$DCG$^M$ &
& 34669.08
& 34603.85
& 34567.22
& 34563.42 \\
 
& A/$^4$DC &
& 34755.89
& 34690.63
& 34567.48
& 34563.68 \\

& A/$^2$DC$^M$ &
& 34755.89
& 34690.63
& 34567.24
& 34563.44 \\

& A/ref.~\citenum{2015_southworth_jcp_xef2} &
& 34752.00
& 34690.90
& 34567.51
& 34563.71 \\

& exp.~\cite{2015_southworth_jcp_xef2} &
& 
& 
& 
&  34565.13 \\
&&&&&&\\

L$_1$
& D/$^2$DC$^M$ &
& 5509.35
& 5473.66
& 5460.49
&  \\

& D/$^2$DCG$^M$ &
& 5501.17
& 5465.51
& 5460.49
&  \\

& exp.~\cite{Oura2019} &
& 
& 
& 
&  5452.7 \\

L$_2$
& D/$^2$DC$^M$ &
& 5161.45
& 5122.38
& 5109.30
&  \\

& D/$^2$DCG$^M$ &
& 5146.77
& 5107.77
& 5109.30
&  \\

& exp.~\cite{Oura2019} &
& 
& 
& 
&  5106.7 \\

L$_3$
& D/$^2$DC$^M$ &
& 4835.59
& 4796.85
& 4787.51
&  \\

& D/$^2$DCG$^M$ &
& 4825.15
& 4786.46
& 4784.51
&  \\

& exp.~\cite{Oura2019} &
& 
& 
& 
&  4786.7 \\
&&&&&&\\

M$_4$
& D/$^2$DC$^M$ &
& 708.13
& 689.63
& 689.01
& 688.31 \\

& D/$^2$DCG$^M$ &
& 707.46
& 688.98
& 689.01
& 688.31 \\

& A/ref.\citenum{2015_southworth_jcp_xef2} &
& 707.5
& 689.6
& 688.98
& 688.28 \\

& exp.~\cite{2015_southworth_jcp_xef2} &
& 
& 
& 
&  689.23 \\

M$_5$
& D/$^2$DC$^M$ &
& 694.90
& 676.71
& 676.32
& 675.62 \\

& D/$^2$DCG$^M$ &
& 694.58
& 676.30
& 676.32
& 675.62 \\

& A/ref.\citenum{2015_southworth_jcp_xef2} &
& 694.6
& 676.7
& 676.08
& 675.38 \\

& exp.~\cite{2015_southworth_jcp_xef2} &
& 
& 
& 
&  676.44 \\

\hline
\multicolumn{7}{c}{XeF$_2$} \\
\hline
K
& D/$^4$DC &
& 34759.79
& 34694.48
& 34567.08
& 34566.18 \\

& D/$^2$DC$^M$ &
& 34759.79
& 34693.30
& 34569.90
& 34566.00 \\

& A/$^2$DC$^M$ &
& 34759.76
& 34694.28
& 34570.89
& 34566.99 \\

& A/ref.~\citenum{2015_southworth_jcp_xef2} &
& 34755.80		
& 34694.50
& 34571.10
& 34567.20\\

& exp.~\cite{2015_southworth_jcp_xef2} &
& 
& 
& 
&  34567.4 \\
&&&&&&\\

M$_4$
& D/$^2$DC$^M$ &
& 711.99
& 693.23 
& 692.61 
& 691.61 \\ 

& A/$^2$DC$^M$ &
& 711.74
& 693.32
& 692.70
& 691.70 \\

& A/ref.~\citenum{2015_southworth_jcp_xef2} &
& 711.9	
& 693.9
& 693.28
& 692.28 \\

& exp.~\cite{1974_carroll_jacs_xef2_exp} &
& 
& 
& 
&  692.09 \\

M$_5$
& D/$^2$DC$^M$ &
& 698.49
& 680.22
& 679.93
& 678.93 \\

& A/$^2$DC$^M$ &
& 698.46
& 680.30
& 680.01
& 679.01 \\

& A/ref.~\citenum{2015_southworth_jcp_xef2} &
& 698.6		
& 680.6
& 680.31
& 679.31 \\

& exp.~\cite{1974_carroll_jacs_xef2_exp} &
& 
& 
& 
&  679.31 \\

\hline
\multicolumn{7}{c}{CH$_3$I}\\
\hline
K
& D/$^2$DCG$^M$ &
& 33278.46
& 33213.79
& 33180.89
&  \\

& exp.~\cite{Boudjemia2019} &
& 
& 
& 
&  33175.20 \\
&&&&&&\\

L$_1$
& D/$^2$DCG$^M$ &
& 5244.04
& 5208.55
& 5203.95
&  \\

& exp.~\cite{Boudjemia2019} &
& 
& 
& 
&  5197.47 \\
\hline
\end{tabular}
\end{table}

The results are shown in Table~\ref{xe-excitations}, which displays selected excitation energies from different edges to the LUMO. We chose these particular transitions since both experimental and computational data are available in the literature,\cite{2015_southworth_jcp_xef2, 2019_southworth_pra_kedge} which can be used as reference values. In particular, the experimental values are tabulated with the theoretical energies obtained in this study.

As already observed for the ionization processes, inclusion of the Gaunt interaction, accounting for the magnetic interaction between the electrons, lowers the excitation energy. This was expected as, in general, the inclusion of this term shifts the orbital energies of the inner core orbitals up (or lowers them, in term of absolute energy) and, at the same time, it reduces the spin-orbit coupling,\cite{2011_saue_cpc_hamprimer} which is also reflected by the results reported herein. Indeed, the difference in the excitation energy from 3$d_{3/2}$ and 3$d_{5/2}$ to LUMO is slightly higher when the $^2$DC$^M$ Hamiltonian is employed.

Upon inclusion of the Gaunt term in the Hamiltonian, the excitation energies differ by at most 0.4\% from the corresponding experimental value, specifically for $1s$(F). This mismatch is higher than the others, which can be due to the fact that the CVS space was reduced so that this was the highest orbital, necessary in order to get this excitation. 

Thus, this space excludes the $3d$ orbitals and in particular $3d_{3/2}$. The next higher disagreement between theory and experiment is found when exciting the $1s$(Xe), although by only 0.1\%. Finally, the best agreement is found for both $3d$ orbitals, with an error of only 0.04\% (3$d_{3/2}$) and 0.02\% (3$d_{5/2}$) from experiment. The computed energies are thus within the experimental error. In order to obtain the excitation from the (3$d_{3/2}$) orbital, the same strategy of excluding all higher orbitals from the CVS space, hence excluding (3$d_{5/2}$), might be the reason of the subtly larger disagreement. 

Nonetheless, these errors are actually quite small. Therefore, we can conclude that the CVS-EOM-CC method implemented in this work gives satisfactory core excitation energies.

\begin{table}[ht]
\centering
\caption{XeF$_2$. Comparison between experimental gas-phase excitation energies and calculations (in eV). The calculated excitation energies have been obtained at the CVS-EOM-CCSD level of theory using the  $^2$DC$^M$ and the $^2$DCG$^M$ Hamiltonians without contributions from the ($SS|SS$) integrals and the Dyall (D) basis set\label{xe-excitations}}
\begin{tabular}{lll p{0.1cm} r}
\hline
Edge
& transition
& model & 
& $E$ \\
\hline
K
& $1s$ (Xe) $\to$ LUMO
& D/$^2$DC$^M$ &
& 34686.70
\\
&
& D/$^2$DCG$^M$ &
&34599.95 \\
&
& exp.\cite{2019_southworth_pra_kedge} &
& 34557.4  \\

& $1s$ (F) $\to$ LUMO
& D/$^2$DC$^M$ &
& 685.91  \\
& 
& D/$^2$DCG$^M$ &
& 685.70 \\
&
& exp.\citenum{2015_southworth_jcp_xef2} &
&  $682.8 \pm 0.3$ \\

M$_4$
& $3d_{3/2}$ $\to$ LUMO
& D/$^2$DC$^M$ &
& 683.71 \\
& 
& D/$^2$DCG$^M$ &
& 683.06 \\
&
& exp.\citenum{2015_southworth_jcp_xef2} &
& $682.8 \pm 0.3$\\

M$_5$
& $3d_{5/2}$ $\to$ LUMO
& D/$^2$DC$^M$ &
& 670.34 \\
& 
& D/$^2$DCG$^M$ &
& 670.04 \\
&
& exp.\citenum{2015_southworth_jcp_xef2} &
& $669.9 \pm 0.3$ \\
\hline
\end{tabular}
\end{table}

\section{Conclusions}\label{sec:conclusions}

We have presented  an implementation of the core-valence separation for the equation-of-motion coupled-cluster method in the \textsc{Dirac} program. This implementation, which is based on a flexible framework for defining projection operators, enables the calculation ionization and excitation energies for all 4-component based Hamiltonians available in \textsc{Dirac}, and consequently for non-relativistic Hamiltonians as well. 
We have applied our implementation to the calculation of core electron binding energies for halogen 
(CH$_3$I, X$^{-}$, and HX, X = Cl--At) and Xe species (Xe, XeF$_2$). For the latter, we also briefly explored the calculation of core excitations. With these systems, we have investigated the performance of different approximations to the original CVS approach, the basis set effects, and the performance of different classes of 2-component approximations.

For highly symmetric species such as the lighter halides, for which we are able to exactly diagonalize the original similarity-transformed Hamiltonian, we show that the CVS energies closely match those from exact diagonalization at all edges. For the overall test set, our assessment of the different approximations indicates the one which more closely matches the performance of the original CVS scheme employs both frozen core and the removal of doubly excited determinants containing only core occupied spinors. Taken individually, these approximations yield sizeable overestimations and underestimations to the core ionization energies, respectively. 

With respect to the Hamiltonians, we observe first that calculations in which the transformation to 2-component is performed after the SCF step ($^2$DC$^M$) are nearly indistinguishable from the equivalent 4-component ones ($^4$DC), though non-negligible discrepancies appear at the K edge of fifth and sixth--row elements. We have traced these discrepancies to the use of the uncorrected 2-electron operator in the 2-component molecular mean-field scheme ($^2$DC$^M$ and $^2$DCG$^M$). More approximate approaches in which the transformation to 2-component is carried out before the SCF step (X2C-AMFI and X2C-G-AMFI), on the other hand, offer at best qualitative accuracy. Second, our results underscore the importance of explicitly considering SSSS-type integrals, in particular from the fifth row onwards.

A comparison to experimental results for Xe, XeF$_2$ and CH$_3$I underscores the importance of the QED, Breit and higher order correlation effects to approach experimental results. While our calculations including the Gaunt interaction recover a significant fraction of the Breit interaction, the gauge term remains quite significant for the K and L edges, and it must be accounted for.
 
In view of these findings, we consider the 4-component based CVS-EOM-CCSD as a reliable approach for investigating core properties throughout the periodic table. Apart from its intrinsic interest, it may serve as a basis for further investigations of the reliability of more approximate schemes for atoms beyond the fifth row, as well as to verify whether such approximations result in significant changes for properties such as transition moments, that require the determination of the excited-state wavefunctions.  


\section{Acknowledgements}

ASPG and LH acknowledge support from PIA ANR project CaPPA (ANR-11-LABX-0005-01), the Franco-German project CompRIXS (Agence nationale de la recherche ANR-19-CE29-0019, Deutsche Forschungsgemeinschaft JA 2329/6-1), I-SITE ULNE projects OVERSEE and MESONM International Associated Laboratory (LAI) (ANR-16-IDEX-0004), the French Ministry of Higher Education and Research, region Hauts de France council and European Regional Development Fund (ERDF) project CPER CLIMIBIO, and the French national supercomputing facilities (grants DARI A0050801859, A0070801859), as well as discussions with Trond Saue at the 2020 \textsc{Dirac} Developers' meeting and following the appearance of the manuscript as a preprint.
MLV and SC acknowledge support from DTU Chemistry (start-up Ph.D. grant),
from the Independent Research Fund Denmark--Natural Sciences, Research Project 2, grant no. 7014-00258B, 
and from the European Union's
Horizon 2020 research and innovation program under the Marie Sk{\l}odowska-Curie European Training Network {\em COmputational Spectroscopy
In Natural sciences and Engineering (COSINE)}, grant
agreement no. 765739. Support from the COST action CM1405 {\em{MOLIM: Molecules in Motion}} (short-term mobility grant to MLV to visit ASPG) is also acknowledged.

\bibliography{ms}

\end{document}


\title{Supplemental Information: Relativistic EOM-CCSD for core-excited and core-ionized state energies based on the 4-component Dirac-Coulomb(-Gaunt) Hamiltonian.}

\author{Lo\"ic Halbert}
\email{loic.halbert@univ-lille.fr}
\affiliation{Universit\'e de Lille, CNRS, 
UMR 8523 -- PhLAM -- Physique des Lasers, 
Atomes et Mol\'ecules, F-59000 Lille, France}

\author{Marta L. Vidal}
\email{malop@kemi.dtu.dk}
\affiliation{Department of Chemistry, Technical University of Denmark, DK-2800 Kongens Lyngby, Denmark}

\author{Avijit Shee}
\email{ashee@umich.edu}
\affiliation{Department of Chemistry, University of Michigan, Ann Arbor, Michigan 48109,USA}

\author{Sonia Coriani}
\email{soco@kemi.dtu.dk}
\affiliation{Department of Chemistry, Technical University of Denmark, DK-2800 Kongens Lyngby, Denmark}

\author{Andr{\'e} Severo Pereira Gomes}
\email{andre.gomes@univ-lille.fr}
\affiliation{Universit\'e de Lille, CNRS, 
UMR 8523 -- PhLAM -- Physique des Lasers, 
Atomes et Mol\'ecules, F-59000 Lille, France}

\begin{abstract}
The datasets, notebooks and figures are available at \url{https://doi.org/10.5281/zenodo.4116367}
\end{abstract}
\maketitle

\begin{figure}[b]
\centering
\includegraphics[width=\textwidth]{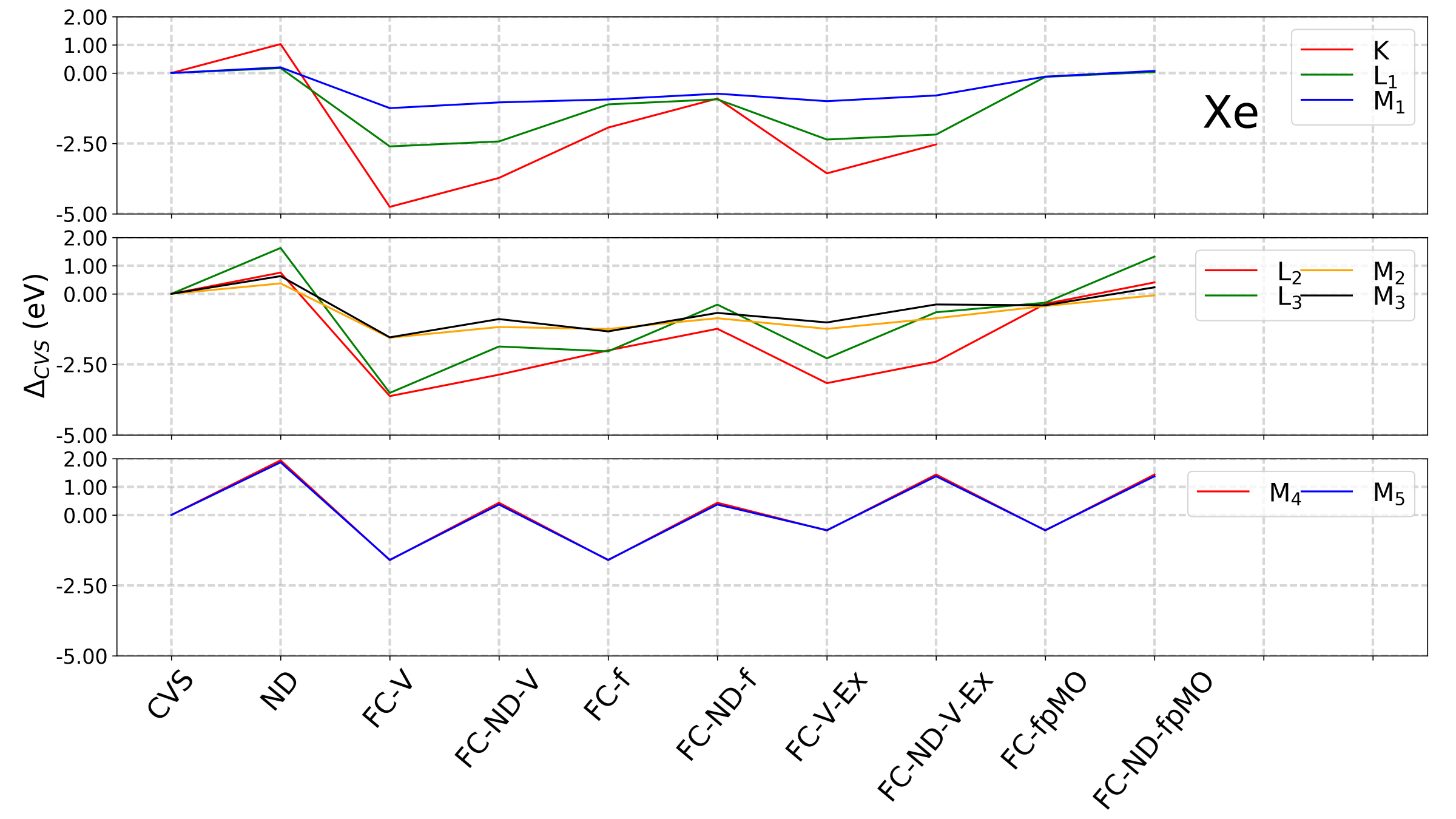}
\caption{Effect of the CVS variants on the different binding energies of Xe. Values (in eV) are relative to the original CVS approach.
\label{fig:Xe}
}
\end{figure}

\begin{figure}[b]
\centering
\includegraphics[width=\textwidth]{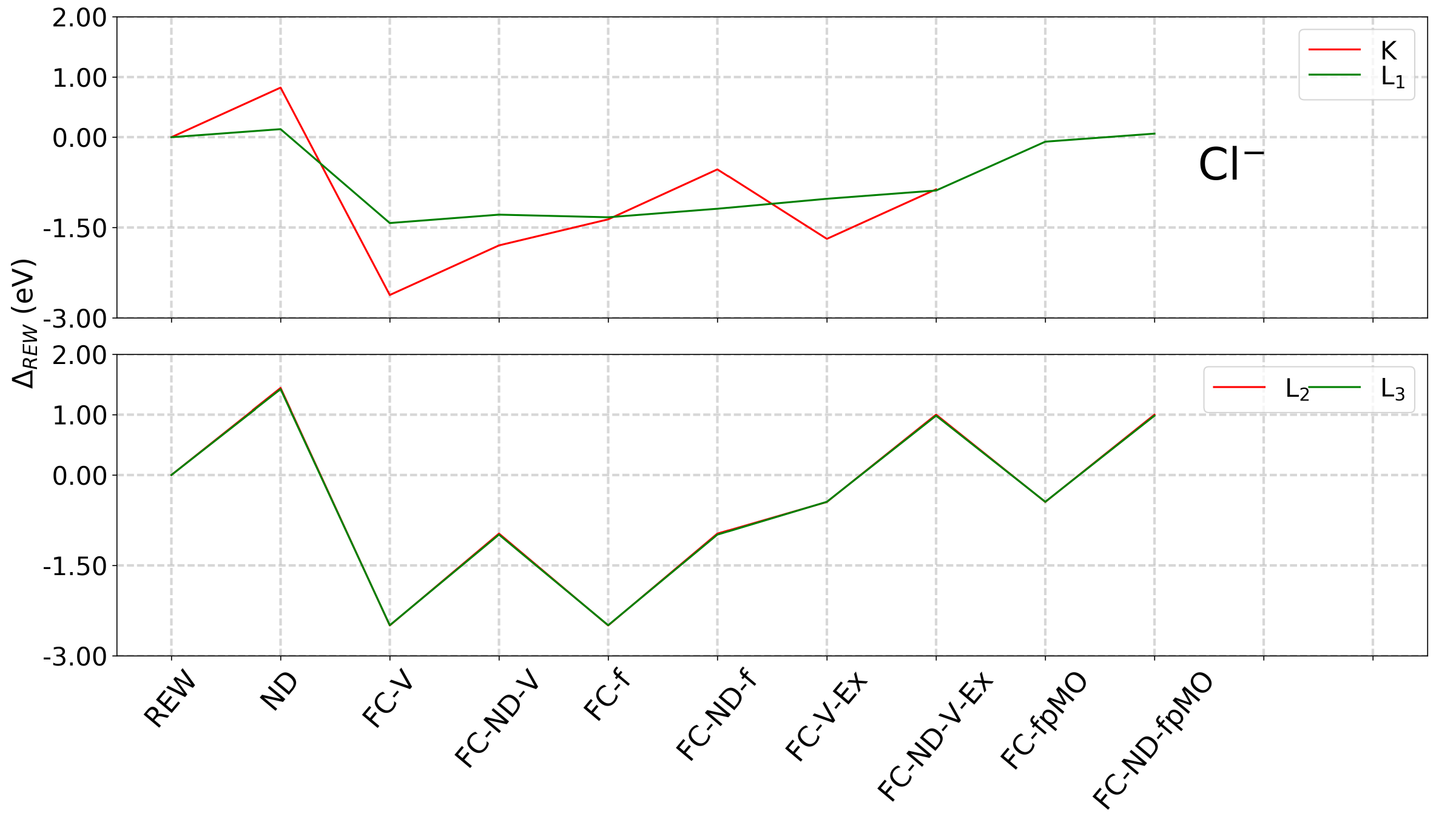}
\caption{Effect of the REW variants on the different binding energies of Cl$^-$.  Values (in eV) are relative to the original REW approach.\label{fig:rew:cl}}
\end{figure}

\begin{figure}[b]
\centering
\includegraphics[width=\textwidth]{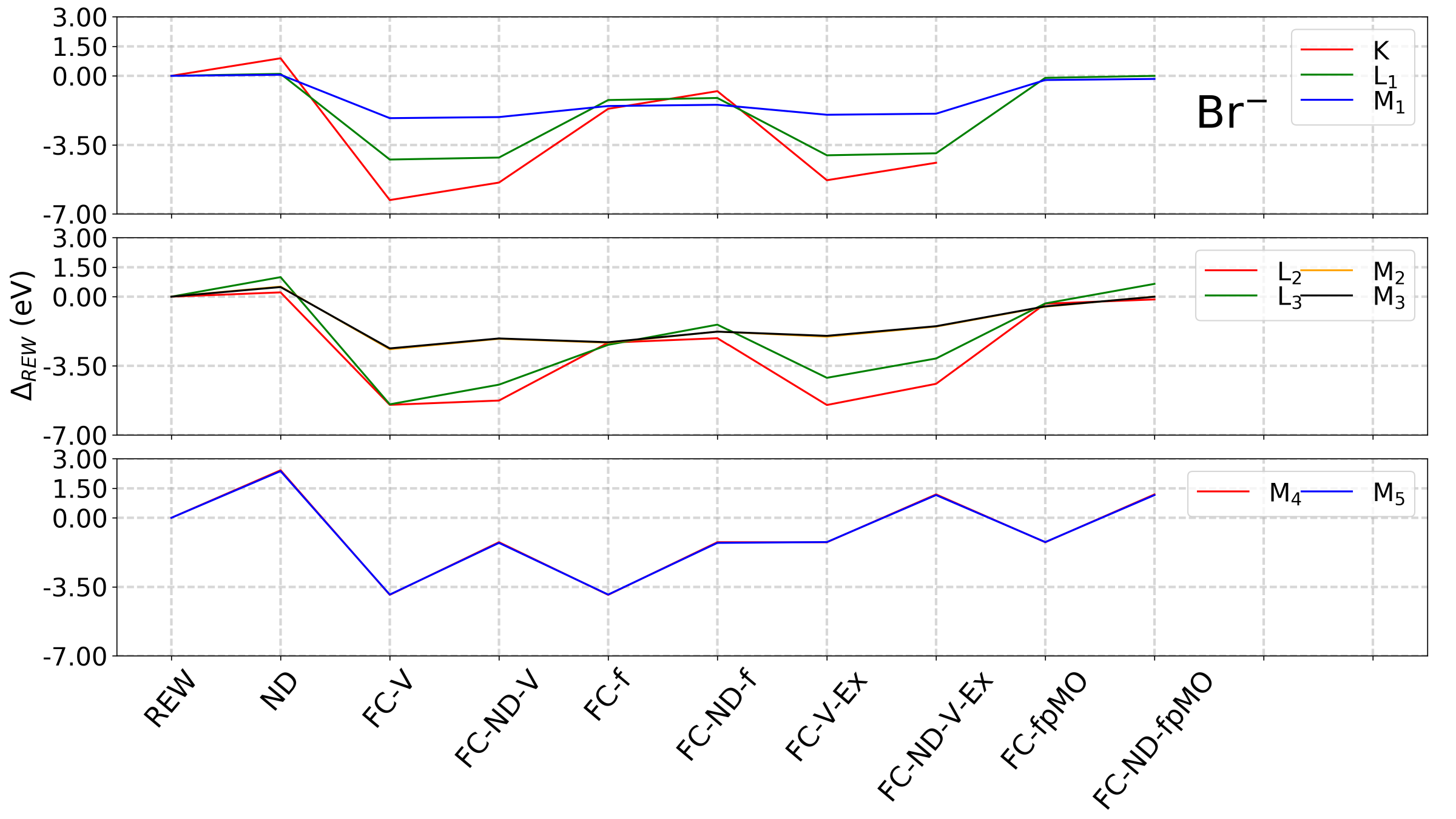}
\caption{Effect of the REW variants on the different binding energies of Br$^-$.  Values (in eV) are relative to the original REW approach.}
\label{fig:rew:br}
\end{figure}

\begin{figure}[b]
\centering
\includegraphics[width=\textwidth]{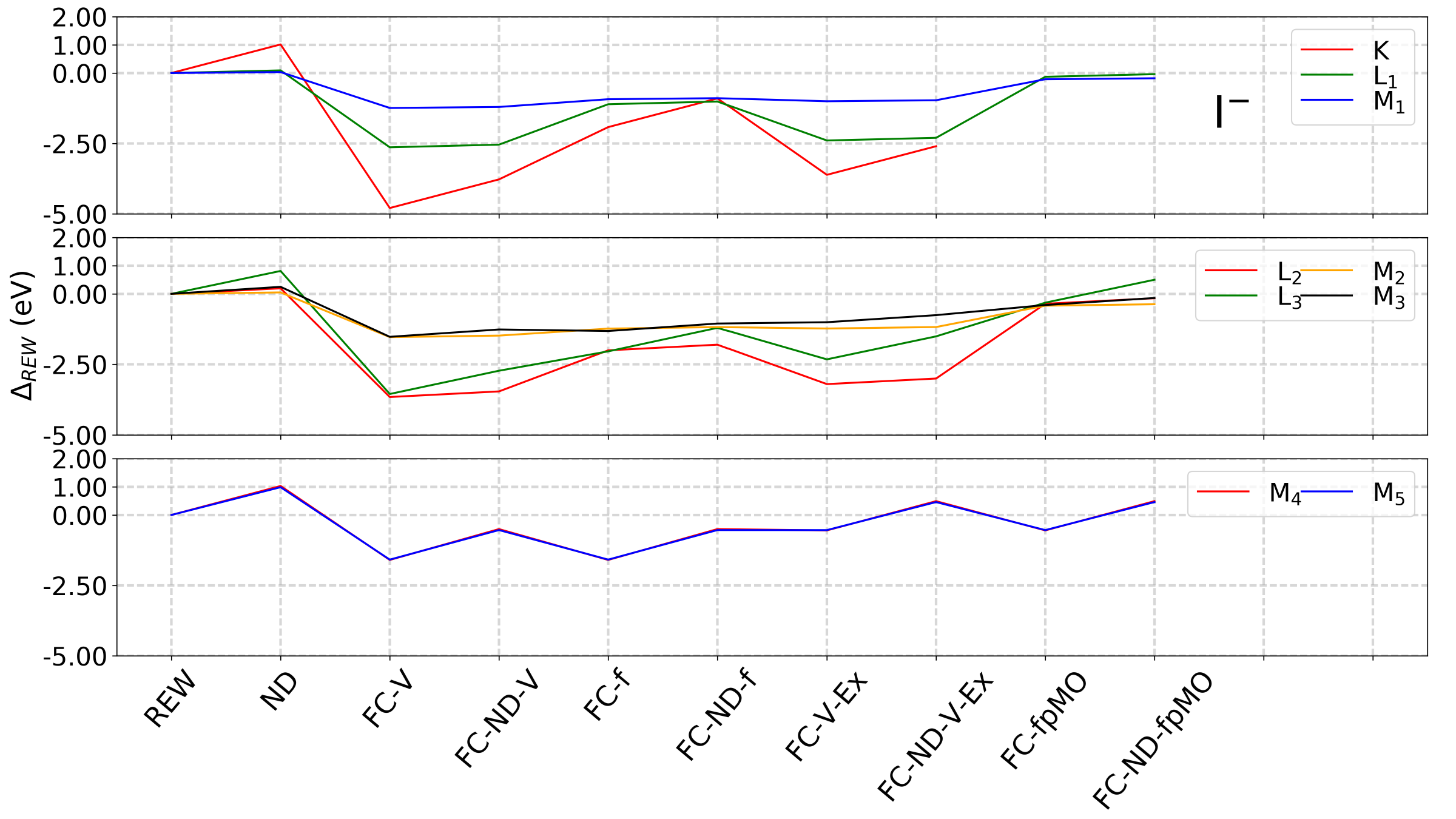}
\caption{Effect of the REW variants on the different binding energies of I$^-$.  Values (in eV) are relative to the original REW approach.}
\label{fig:rew:i}
\end{figure}

\begin{figure}[b]
\centering
\includegraphics[width=\textwidth]{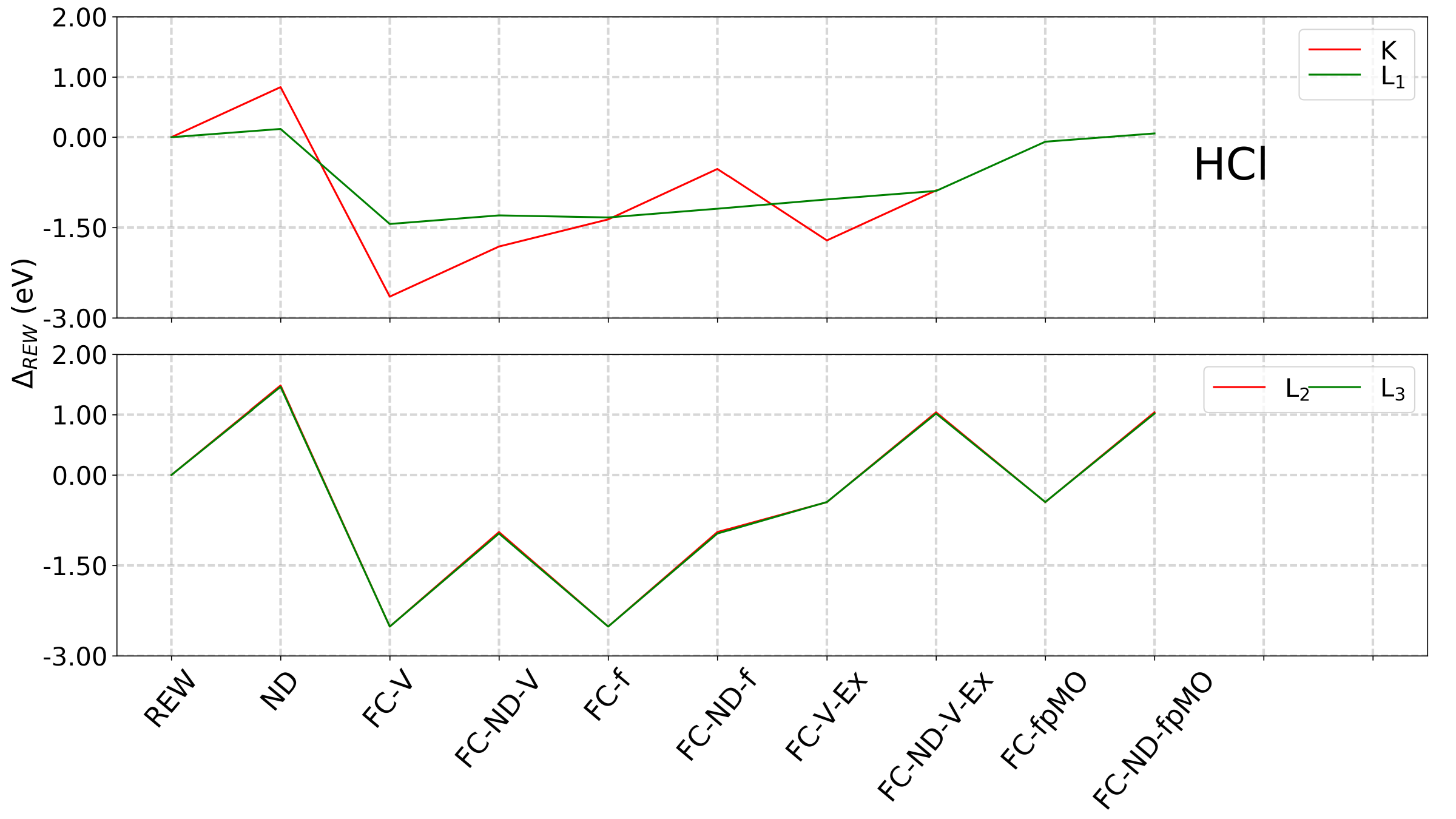}
\caption{Effect of the REW variants on the different binding energies of HCl.  Values (in eV) are relative to the original REW approach.}
\label{fig:rew:hcl}
\end{figure}

\begin{figure}[b]
\centering
\includegraphics[width=\textwidth]{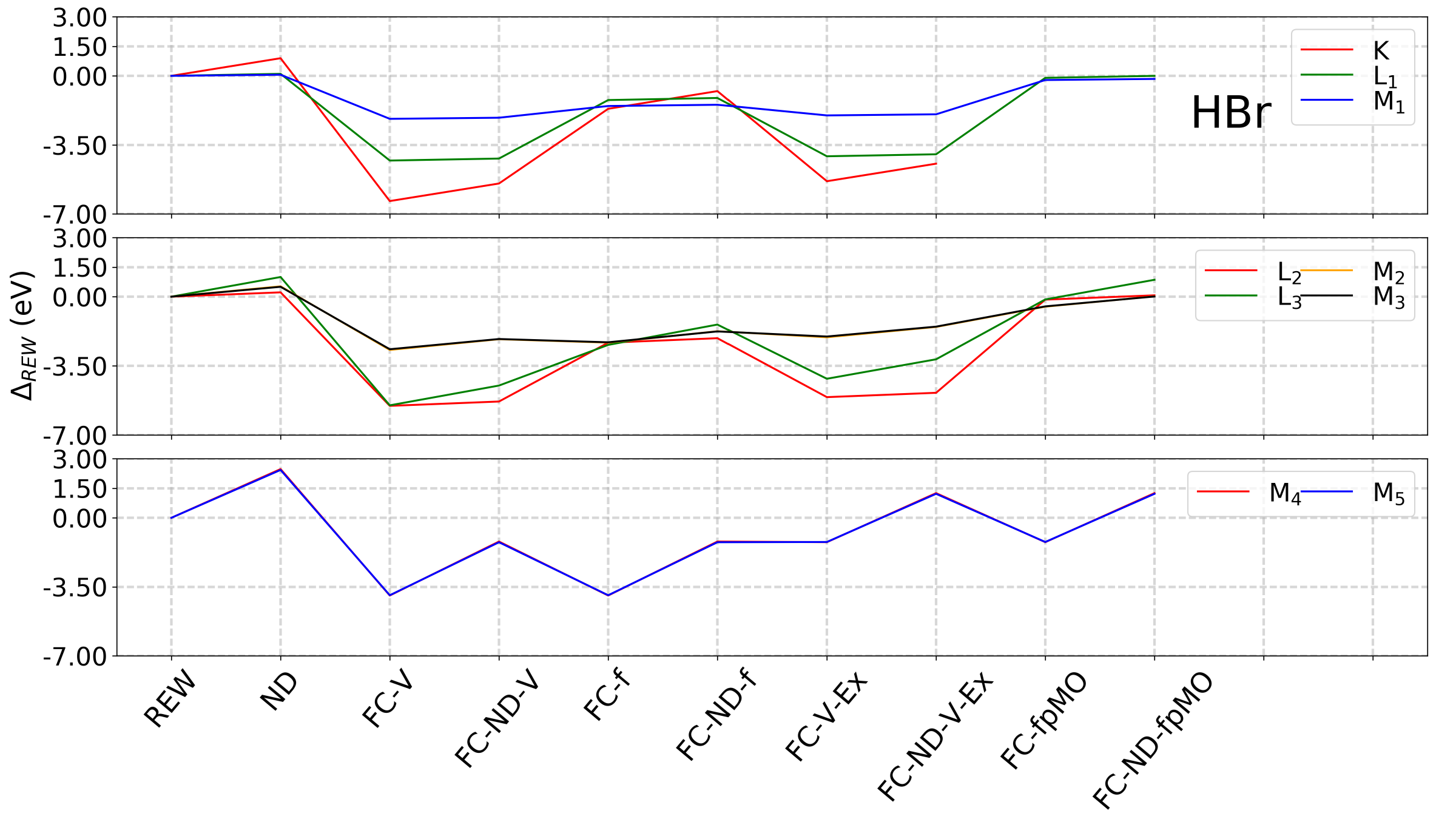}
\caption{Effect of the REW variants on the different binding energies of HBr.  Values (in eV) are relative to the original REW approach.}
\label{fig:rew:hbr}
\end{figure}

\begin{figure}[b]
\centering
\includegraphics[width=\textwidth]{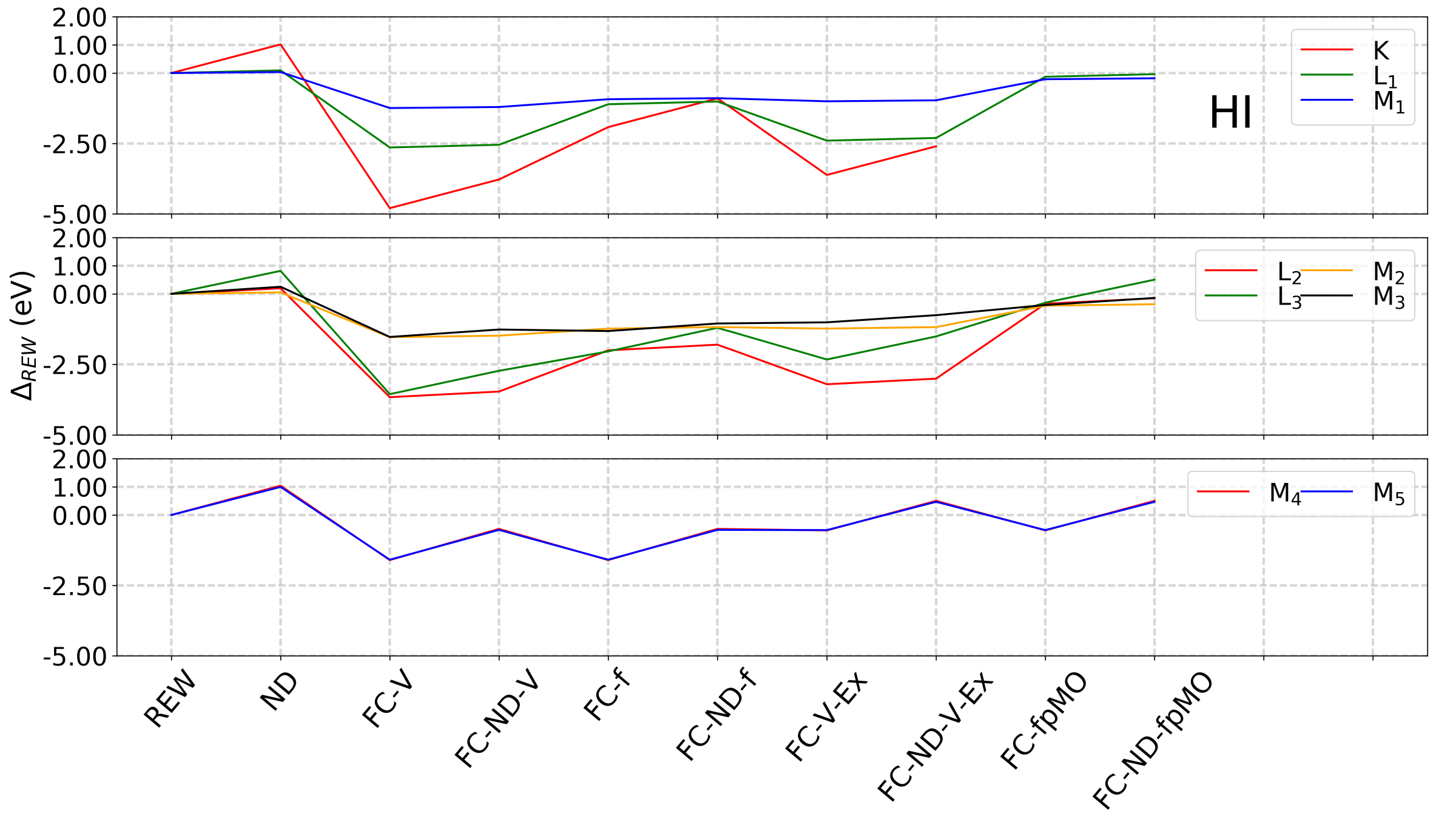}
\caption{Effect of the REW variants on the different binding energies of HI.  Values (in eV) are relative to the original REW approach.}
\label{fig:rew:hi}
\end{figure}

\clearpage

\begin{table*}[p]
    \centering
\caption{Operation counts for the construction of a single $\sigma$ vector for EOM-EE and EOM-IP wavefunctions for different variants (P: CVS-EOM using projectors; A: analytical CVS-EOM; A(FC-f) : analytical CVS-EOM with the frozen core approximation; A(ND): analytical CVS-EOM with the ND approximation; A(FC-f + ND): analytical CVS-EOM combining FC-f and ND; see manuscript for additional details). The operation counts are calculated employing the number of virtual (Cl$^-$: 208; At$^-$: 504 with dyall.acv3z basis sets) valence occupied and core spinors for each edge (with 2, 8, 18 and 60 core spinors for the K, L$_3$, M$_5$ and N$_7$ edges, respectively) from the working equations from Shee et al. \emph{J. Chem.\ Phys.} (2018) \textbf{149}, 174113. For convenience these expressions are reproduced below, with different colors identifying the terms that relate to the A(FC-f), A(ND) and A(FC-f + ND) variants.}
    \label{tab:operation_count}
\begin{tabular}{ll ccccc }
       \hline
       &      & \multicolumn{5}{c}{CVS variants} \\
       \cline{3-7}
System & Edge & P & A & A(FC-f) & A(ND) & A(FC-f + ND)     \\
\hline 
\multicolumn{7}{c}{EOM-EE} \\
\hline
Cl$^-$  & K      & 2.0E+14 & 1.4E+13 & 8.6E+12 & 1.1E+13 & 7.7E+12 \\
        & L$_3$  & 2.0E+14 & 1.1E+14 & 1.1E+13 & 3.2E+13 & 4.9E+12 \\
        &&&&&&\\
At$^-$  & K      & 3.6E+18 & 4.4E+16 & 3.9E+16 & 4.1E+16 & 3.8E+16 \\
        & L$_3$  & 3.6E+18 & 2.7E+17 & 1.6E+17 & 2.0E+17 & 1.4E+17 \\
        & M$_5$  & 3.6E+18 & 1.0E+18 & 2.6E+17 & 4.8E+17 & 1.8E+17 \\
        & N$_7$  & 3.6E+18 & 2.6E+18 & 1.1E+17 & 4.6E+17 & 3.4E+16 \\
\hline
\multicolumn{7}{c}{EOM-IP} \\
\hline  
Cl$^-$  & K      & 5.2E+12 & 1.1E+10 & 3.3E+07 & 9.2E+09 & 2.7E+07 \\
        & L$_3$  & 5.2E+12 & 2.0E+11 & 1.9E+08 & 5.8E+10 & 6.8E+07 \\
        &&&&&&\\
At$^-$  & K      & 2.0E+16 & 3.7E+12 & 3.9E+09 & 3.6E+12 & 3.7E+09 \\
        & L$_3$  & 2.0E+16 & 8.9E+13 & 2.1E+10 & 7.4E+13 & 1.7E+10 \\
        & M$_5$  & 2.0E+16 & 6.2E+14 & 6.7E+10 & 3.4E+14 & 3.7E+10 \\
        & N$_7$  & 2.0E+16 & 2.2E+15 & 1.5E+11 & 3.1E+14 & 3.6E+10 \\
\hline 
\end{tabular}
\end{table*}

\clearpage

Below are the equations for EOM-IP and EOM-EE right-hand side $\sigma$-vector equations (we employ $I,J, \ldots$ to denote occupied core orbitals and $i_v, j_v, \ldots$ occupied valence orbitals), color-coded as to indicate which terms would be present or absent for different approimations:\label{work_eqs}
\begin{itemize}
 \item Black: always calculated
 \item {\color{blue}{Blue: calculated if ND is not invoked}}
 \item {\color{green}{Green: calculated if FC-f is not invoked}}
 \item {\color{red}{Red: calculated if neither FC-f nor ND are invoked}}
\end{itemize}

\begin{itemize}
\item[EE:] 
\begin{align}
^R \sigma_I^a = {} & \sum_e \overline{F}^a_e r^e_I 
                   - \sum_M \overline {F}^M_I  r_M^a 
                   + \sum_{m_ve} \overline{F}^{m_v}_e r^{ea}_{m_vI}
{\color{blue}{     + \sum_{Me} \overline{F}^M_e r^{ea}_{MI} }}
                   + \sum_{Me} W^{Ma}_{eI} r^e_M \nonumber  \\ & 
                   + \sum_{m_v,e>f} W^{am_v}_{ef} r^{ef}_{Im_v} 
{\color{blue}{     + \sum_{M,e>f} W^{aM}_{ef} r^{ef}_{IM}}} 
                   - \sum_{Mn_v,e}  W^{Mn_v}_{Ie} r_{Mn_v}^{ae} 
{\color{blue}{     - \sum_{M>N,e}  W^{MN}_{Ie} r_{MN}^{ae}}} \nonumber \\ &
\end{align}
\break
\begin{align}
^R \sigma_{Ij_v}^{ab} =& - P(ab) \sum_M W^{Mb}_{Ij_v} r^a_M 
                       + \sum_e W^{ab}_{ej_v} r^e_I
{\color{green}{        + P(ab) \sum_{efM} (W^{bM}_{fe} r^e_M) t^{af}_{Ij_v} }} \nonumber \\ &
{\color{green}{        + \sum_{Mn_ve} (W^{n_vM}_{j_ve} r^e_M) t^{ab}_{In_v} }}
                       -\sum_{Mn_ve} (W^{n_vM}_{Ie} r^e_M) t^{ab}_{j_vn_v}  
{\color{green}{        + \sum_{MNe} (W^{NM}_{j_ve} r^e_M) t^{ab}_{IN} }} 
{\color{green}{        - \sum_{MNe} (W^{NM}_{Ie} r^e_M) t^{ab}_{j_vN} }}  \nonumber \\ &
                       + P(ab) \sum_e \overline{F}^b_e r^{ae}_{Ij_v} 
                       - \sum_{m_v} \overline{F}^{m_v}_{j_v} r^{ab}_{Im_v} 
{\color{blue}{         - \sum_{M} \overline{F}^{M}_{j_v} r^{ab}_{IM} }}
                       + \sum_{M} \overline{F}^{M}_{I} r^{ab}_{j_vM} 
                       + \sum_{Mn_v} W^{Mn_v}_{Ij_v} r ^{ab}_{Mn_v} \nonumber \\ &
{\color{blue}{         + \sum_{M>N} W^{MN}_{Ij_v} r ^{ab}_{MN} }} 
                       + P(ab) \sum_{m_ve} W^{m_vb}_{ej_v} r^{ae}_{Im_v} 
{\color{blue}{         + P(ab) \sum_{Me} W^{Mb}_{ej_v} r^{ae}_{IM} }} \nonumber \\ &
                       - P(ab) \sum_{Me} W^{Mb}_{eI} r^{ae}_{j_vM} 
{\color{green}{        - P(ab) \sum_{ef,Mn_v} (V^{n_vM}_{fe} r^{ea}_{Mn_v}) t^{fb}_{Ij_v} }} 
{\color{red}{        - P(ab) \sum_{ef,M>N} (V^{NM}_{fe} r^{ea}_{MN}) t^{fb}_{Ij_v} }} \nonumber \\ &
                       - P(Ij_v) \sum_{e>f, m_vn_v} (V^{n_vm_v}_{fe} r ^{fe}_{Im_v}) t^{ba}_{j_vn_v} 
{\color{blue}{         - P(Ij_v) \sum_{e>f, Mn_v} (V^{n_vM}_{fe} r ^{fe}_{IM}) t^{ba}_{j_vn_v} }} \nonumber \\ &
{\color{green}{        + \sum_{e>f, Mn_v} (V^{n_vM}_{fe} r ^{fe}_{j_vM}) t^{ba}_{In_v} }}
{\color{red}{          - \sum_{e>f, MN} (V^{NM}_{fe} r ^{fe}_{IM}) t^{ba}_{j_vN}  }}  
{\color{green}{        + \sum_{e>f, MN} (V^{NM}_{fe} r ^{fe}_{j_vM}) t^{ba}_{IN} }}   \nonumber \\ &
{\color{red}{          - \sum_{e>f, m_vN} (V^{Nm_v}_{fe} r ^{fe}_{Im_v}) t^{ba}_{j_vN}  }}           
                       + \sum_{e>f} W^{ab}_{ef} r^{ef}_{Ij_v} \nonumber\\                      
\end{align}
{\color{blue}{     
\begin{align}
^R \sigma_{IJ}^{ab} =& - P(ab) \sum_M W^{Mb}_{IJ} r^a_M 
                       + P(IJ) \sum_e W^{ab}_{eJ} r^e_I 
{\color{red}{          + P(ab) \sum_{efM} (W^{bM}_{fe} r^e_M) t^{af}_{IJ}  }} \nonumber \\ & 
{\color{red}{          + P(IJ) \sum_{Mn_ve} (W^{n_vM}_{Je} r^e_M) t^{ab}_{In_v} }}
{\color{red}{          + P(IJ) \sum_{MNe} (W^{NM}_{Je} r^e_M) t^{ab}_{IN} }}
                       + P(ab) \sum_e \overline{F}^b_e r^{ae}_{IJ}  \nonumber \\ &
                       - P(IJ) \sum_{m_v} \overline{F}^{m_v}_J r^{ab}_{Im_v}  
                       - P(IJ) \sum_{M} \overline{F}^{M}_J r^{ab}_{IM}
                       + \sum_{Mn_v} W^{Mn_v}_{IJ} r ^{ab}_{Mn_v}
                       + \sum_{M>N} W^{MN}_{IJ} r ^{ab}_{MN} \nonumber\\&
                       + P(ab)P(IJ) \sum_{m_ve} W^{m_vb}_{eJ} r^{ae}_{Im_v} 
                       + P(ab)P(IJ) \sum_{Me} W^{Mb}_{eJ} r^{ae}_{IM} 
{\color{red}{          - P(ab) \sum_{ef,Mn_v} (V^{n_vM}_{fe} r^{ea}_{Mn_v}) t^{fb}_{IJ} }}  \nonumber \\&
{\color{red}{          - P(ab) \sum_{ef,M>N} (V^{NM}_{fe} r^{ea}_{MN}) t^{fb}_{IJ} }}
{\color{red}{          - P(IJ) \sum_{e>f, m_vn_v} (V^{n_vm_v}_{fe} r ^{fe}_{Im_v}) t^{ba}_{Jn_v} }} 
{\color{red}{          - P(IJ) \sum_{e>f, Mn_v} (V^{n_vM}_{fe} r ^{fe}_{IM}) t^{ba}_{Jn_v}  }} \nonumber\\&
{\color{red}{          - P(IJ) \sum_{e>f, m_vN} (V^{Nm_v}_{fe} r ^{fe}_{Im_v}) t^{ba}_{JN} }} 
{\color{red}{          - P(IJ) \sum_{e>f, MN} (V^{NM}_{fe} r ^{fe}_{IM}) t^{ba}_{JN} }}
                        + \sum_{e>f} W^{ab}_{ef} r^{ef}_{IJ}                         
\end{align}
}}

\item [IP:] \begin{align}
^R \sigma_I = {} & - \sum_M \overline {F}^M_I  r_M 
                   + \sum_{m_ve} \overline{F}^{m_v}_e  r_{Im_v}^{e} 
{\color{blue}{     + \sum_{Me} \overline{F}^{M}_e  r_{IM}^{e} }} 
                   + \sum_{Mn_v,e}  W^{Mn_v}_{Ie} r_{Mn_v}^{e} 
{\color{blue}{     + \sum_{M>N,e}  W^{MN}_{Ie} r_{MN}^{e} }}
\end{align}
\begin{align}
^R \sigma_{Ij_v}^{a} = {} & - \sum_M W^{Ma}_{Ij_v}  r_M 
                     - \sum_M \overline{F}^m_I r^a_{Mj_v}
                     + \sum_{m_v} \overline{F}^{m_v}_{j_v} r^a_{{m_v}I}
{\color{blue}{       + \sum_M \overline{F}^M_{j_v} r^a_{MI} }} \nonumber \\
                {} & + \sum_{Mn_v} W ^{Mn_v}_{Ij_v} r_{Mn_v}^a
{\color{blue}{       + \sum_{M>N} W ^{MN}_{Ij_v} r_{MN}^a }} \nonumber \\ &
                     - \sum_{Me} W^{Ma}_{Ie} r_{Mj_v}^e
                     + \sum_{m_ve} W^{m_va}_{j_ve} r_{m_vI}^e 
{\color{blue}{       + \sum_{Me} W^{Ma}_{j_ve} r_{MI}^e  }} \nonumber \\
               {} &  + \sum_e \overline{F}^a_e r_{Ij_v}^e 
{\color{green}{      + \sum_{Mn_v,ef} (V^{Mn_v}_{ef} r_{Mn_v}^{f}) t^{ea}_{Ij_v} }}                 
{\color{red}{        + \sum_{M>N,ef} (V^{MN}_{ef} r_{MN}^{f}) t^{ea}_{Ij_v} }}                 
\end{align}
{\color{blue}{
\begin{align}
^R \sigma_{IJ}^{a} = {} & - \sum_M W^{Ma}_{IJ}  r_M 
                          - P(IJ) \sum_{m_v} \overline{F}^{m_v}_I r^a_{m_vJ}
                          - P(IJ) \sum_M \overline{F}^M_I r^a_{MJ} \nonumber \\
                     {} & + \sum_{m_vN} W ^{m_vN}_{IJ} r_{m_vN}^a 
                          + \sum_{M>N} W ^{MN}_{IJ} r_{MN}^a
                          - P(IJ) \sum_{m_ve} W^{m_va}_{Ie} r_{m_vJ}^e \nonumber \\
                  {} &    - P(IJ) \sum_{Me} W^{Ma}_{Ie} r_{MJ}^e 
                          + \sum_e \overline{F}^a_e r_{IJ}^e 
{\color{red}{             + \sum_{Mn_v,ef} (V^{mn}_{ef} r_{Mn_v}^{f}) t^{ea}_{IJ} }}
{\color{red}{             + \sum_{M>N,ef} (V^{MN}_{ef} r_{MN}^{f}) t^{ea}_{IJ} }}                 
\end{align}
}}

\end{itemize}

The $\overline{F}$, $W$ and $G$ intermediates are defined as follows:

\begin{align}\label{eq:f_bar_oo}
\overline{F}^{{i_v}}_{{m_v}} = & f^{{i_v}}_{{m_v}} 
                     + \sum_e f^{i_v}_e t^e_{m_v}  
                     + \sum_{en_v} V^{{i_v}n_v}_{{m_v}e} t ^e_{n_v} 
{\color{green}{      + \sum_{eN} V^{{i_v}N}_{{m_v}e} t ^e_N }} 
                     + \sum_{n_v,e>f} V^{{i_v}n_v}_{ef} \tau^{ef}_{{m_v}n_v} 
{\color{green}{      + \sum_{N,e>f} V^{{i_v}N}_{ef} \tau^{ef}_{{m_v}N} }} 
\end{align}
\begin{align}\label{eq:f_bar_oc}
\overline{F}^{{i_v}}_{M} = & f^{{i_v}}_{M} 
{\color{green}{      + \sum_e f^{i_v}_e t^e_M }} 
                     + \sum_{en_v} V^{{i_v}n_v}_{Me} t ^e_{n_v} 
{\color{green}{      + \sum_{eN} V^{{i_v}N}_{Me} t ^e_N }} 
{\color{green}{      + \sum_{n_v,e>f} V^{{i_v}n_v}_{ef} \tau^{ef}_{Mn_v} }}
{\color{green}{      + \sum_{N,e>f} V^{{i_v}N}_{ef} \tau^{ef}_{MN} }} 
\end{align}
\begin{align}\label{eq:f_bar_co}
\overline{F}^{I}_{{m_v}} = & f^{I}_{{m_v}} 
                     + \sum_e f^I_e t^e_{m_v}
                     + \sum_{en_v} V^{In_v}_{{m_v}e} t ^e_{n_v} 
{\color{green}{      + \sum_{eN} V^{IN}_{{m_v}e} t ^e_N }} 
                     + \sum_{n_v,e>f} V^{In_v}_{ef} \tau^{ef}_{{m_v}n_v} 
{\color{green}{      + \sum_{N,e>f} V^{IN}_{ef} \tau^{ef}_{{m_v}N} }} 
\end{align}
\begin{align}\label{eq:f_bar_cc}
\overline{F}^{I}_{M} = & f^{I}_{M} 
{\color{green}{      + \sum_e f^I_e t^e_M }} 
                     + \sum_{en_v} V^{In_v}_{Me} t ^e_{n_v} 
{\color{green}{      + \sum_{eN} V^{IN}_{Me} t ^e_N }} 
{\color{green}{      + \sum_{n_v,e>f} V^{In_v}_{ef} \tau^{ef}_{Mn_v} }}
{\color{green}{      + \sum_{N,e>f} V^{IN}_{ef} \tau^{ef}_{MN} }} 
\end{align}

\begin{align}\label{eq:f_bar_vv}
\overline{F}^{e}_{a} =& f^{e}_{a} 
                     - \sum _{m_v} f^{m_v}_a t_{m_v}^e 
{\color{green}{      - \sum _M f^M_a t_M^e }} 
                     - \sum_{m_vf} V^{m_ve}_{fa} t_{m_v}^f
{\color{green}{      - \sum_{Mf} V^{Me}_{fa} t_M^f }}
                     - \sum_{m_v>n_v,f} V^{m_vn_v}_{af} \tau^{ef}_{m_vn_v} \nonumber \\ &
{\color{green}{      - \sum_{Mn_v,f} V^{Mn_v}_{af} \tau^{ef}_{Mn_v} }}
{\color{green}{      - \sum_{m_vN,f} V^{m_vN}_{af} \tau^{ef}_{m_vN} }}
{\color{green}{      - \sum_{M>N,f} V^{MN}_{af} \tau^{ef}_{MN}}}
\end{align}
\begin{align}\label{eq:f_bar_ov}
\overline{F}^{m_v}_{e} = & f^{m_v}_e 
                       + \sum_{n_vf} V^{m_vn_v}_{ef} t^f_{n_v}
{\color{green}{        + \sum_{Nf} V^{m_vN}_{ef} t^f_N }}
\end{align}
\begin{align}\label{eq:f_bar_cv}
\overline{F}^{M}_{e} = & f^{M}_e 
                       + \sum_{n_vf} V^{Mn_v}_{ef} t^f_{n_v}
{\color{green}{        + \sum_{Nf} V^{MN}_{ef} t^f_N }},
\end{align}
\begin{align}\label{eq:W_cccc}
   W^{IJ}_{MN} = {} & V^ {IJ}_{MN} 
{\color{green}{  + P (MN) \sum_e V^ {IJ}_{eN} t_M^e }}
{\color{green}{  + \sum_{e>f} V^{IJ}_{ef} \tau^ {ef}_{MN} }} 
\end{align}
\begin{align}\label{eq:W_ccco}
   W^{IJ}_{Mn_v} = {} & V^ {IJ}_{Mn_v} 
{\color{green}{  + \sum_e V^ {IJ}_{en_v} t_M^e }}
                 - \sum_e V^ {IJ}_{eM} t_{n_v}^e 
{\color{green}{  + \sum_{e>f} V^{IJ}_{ef} \tau^ {ef}_{Mn_v} }} 
\end{align}
\begin{align}\label{eq:W_coco}
   W^{I{j_v}}_{Mn_v} = {} & V^ {I{j_v}}_{Mn_v} 
{\color{green}{  + \sum_e V^ {I{j_v}}_{en_v} t_M^e }}
                 - \sum_e V^ {I{j_v}}_{eM} t_{n_v}^e 
{\color{green}{  + \sum_{e>f} V^{I{j_v}}_{ef} \tau^ {ef}_{Mn_v} }} 
\end{align}
\begin{align}\label{eq:W_cocc}
   W^{I{j_v}}_{MN} = {} & V^ {I{j_v}}_{MN} 
{\color{green}{  + P (MN) \sum_e V^ {I{j_v}}_{eN} t_M^e }}
{\color{green}{  + \sum_{e>f} V^{I{j_v}}_{ef} \tau^ {ef}_{MN} }} 
\end{align}
\begin{align}\label{eq:W_ovvo}
    W^{{m_v}b}_{e{j_v}} = & V^ {{m_v}b}_{e{j_v}} 
                + {\sum_f V^{{m_v}b}_{ef} t_{j_v}^f}  
                - \sum_{n_v} V^{{m_v}n_v}_{e{j_v}} t_{n_v}^b
{\color{green}{ - \sum_{N} V^{{m_v}N}_{e{j_v}} t_N^b }}\nonumber \\&
                + \sum_{n_vf} V ^ {{m_v}n_v}_{ef} (t^{fb}_{{j_v}n_v} - t^f_{j_v} t^b_{n_v}) 
{\color{green}{ + \sum_{Nf} V ^ {{m_v}N}_{ef} (t^{fb}_{{j_v}N} - t^f_{j_v} t^b_N) }} 
\end{align}
\begin{align}\label{eq:W_ovvc}
    W^{{m_v}b}_{eJ} = & V^ {{m_v}b}_{eJ} 
{\color{green}{ + {\sum_f V^{{m_v}b}_{ef} t_J^f} }} 
                - \sum_{n_v} V^{{m_v}n_v}_{eJ} t_{n_v}^b
{\color{green}{ - \sum_{N} V^{{m_v}N}_{eJ} t_N^b }}\nonumber \\&
{\color{green}{ + \sum_{n_vf} V ^ {{m_v}n_v}_{ef} (t^{fb}_{Jn_v} - t^f_J t^b_{n_v}) }}
{\color{green}{ + \sum_{Nf} V ^ {{m_v}N}_{ef} (t^{fb}_{JN} - t^f_J t^b_N) }} 
\end{align}
\begin{align}\label{eq:W_cvvo}
    W^{Mb}_{e{j_v}} = & V^ {Mb}_{e{j_v}} 
                + {\sum_f V^{Mb}_{ef} t_{j_v}^f}  
                - \sum_{n_v} V^{Mn_v}_{e{j_v}} t_{n_v}^b
{\color{green}{ - \sum_{N} V^{MN}_{e{j_v}} t_N^b }}\nonumber \\&
                 + \sum_{n_vf} V ^ {Mn_v}_{ef} (t^{fb}_{{j_v}n_v} - t^f_{j_v} t^b_{n_v})
{\color{green}{ + \sum_{Nf} V ^ {MN}_{ef} (t^{fb}_{{j_v}N} - t^f_{j_v} t^b_N) }} 
\end{align}
\begin{align}\label{eq:W_cvvc}
    W^{Mb}_{eJ} = & V^ {Mb}_{eJ} 
{\color{green}{ + {\sum_f V^{Mb}_{ef} t_J^f} }} 
                - \sum_{n_v} V^{Mn_v}_{eJ} t_{n_v}^b
{\color{green}{ - \sum_{N} V^{MN}_{eJ} t_N^b }}\nonumber \\&
{\color{green}{ + \sum_{n_vf} V ^ {Mn_v}_{ef} (t^{fb}_{Jn_v} - t^f_J t^b_{n_v}) }}
{\color{green}{ + \sum_{Nf} V ^ {MN}_{ef} (t^{fb}_{JN} - t^f_J t^b_N) }} 
\end{align}
\begin{align}\label{eq:W_vovv}
   W^{am_v}_{ef} = {} & V^{am_v}_{ef} 
                 - \sum_{n_v} V^{n_vm_v}_{ef} t^a_{n_v}
{\color{green}{  - \sum_N V^{Nm_v}_{ef} t^a_N }} 
\end{align}
\begin{align}\label{eq:W_vcvv}
   W^{aM}_{ef} = {} & V^{aM}_{ef} 
               - \sum_{n_v} V^{n_vM}_{ef} t^a_{n_v} 
{\color{green}{- \sum_N V^{NM}_{ef} t^a_N }} 
\end{align}
\begin{align}\label{eq:W_cocv}
   W^{Mn_v}_{Ie} = {} & V^{Mn_v}_{Ie} 
{\color{green}{  + \sum_f t^f_I V^{Mn_v}_{fe} }} 
\end{align}
\begin{align}\label{eq:W_cccv}
   W^{MN}_{Ie} = {} & V^{MN}_{Ie} 
{\color{green}{+ \sum_f t^f_I V^{MN}_{fe} }} 
\end{align}
\begin{align}\label{eq:W_occv}
   W^{m_vN}_{Ie} = {} & V^{m_vN}_{Ie} 
{\color{green}{  + \sum_f t^f_{I} V^{m_vN}_{fe} }} 
\end{align}
\begin{align}\label{eq:W_ocov}
   W^{m_vN}_{i_ve} = {} & V^{m_vN}_{i_ve} 
                   + \sum_f t^f_{i_v} V^{m_vN}_{fe} 
\end{align}
\begin{align}\label{eq:W_ccov}
   W^{MN}_{i_ve} = {} & V^{MN}_{i_ve} 
{\color{green}{+ \sum_f t^f_{i_v} V^{MN}_{fe} }} 
\end{align}
\begin{align}\label{eq:W_cvco}
   W^{Ie}_{Mn_v} = {} & V^{Ie}_{Mn_v} 
{\color{green}{  + \sum_{f} \overline{F}^{I}_{f} t^{ef}_{Mn_v} }} 
                 - \sum_{o_v} W^{Io_v}_{Mn_v} t^e_{o_v}
{\color{green}{  - \sum_O W^{IO}_{Mn_v} t^e_O }} 
{\color{green}{  +  {\sum_{f>g} V^{Ie}_{fg} \tau ^{fg}_{Mn_v}} }}
{\color{green}{  + \sum_f \overline{W}^{Ie}_{fn_v} t^f_M }}\nonumber \\&
                 - \sum_f \overline{W}^{Ie}_{fM} t^f_{n_v} 
                 + { \sum_{o_vf} V^{Io_v}_{Mf} t^{ef}_{n_vo_v}}
{\color{green}{  - \sum_{o_vf} V^{Io_v}_{n_vf} t^{ef}_{Mo_v} }}
{\color{green}{  + { \sum_{Of} V^{IO}_{Mf} t^{ef}_{n_vO}} }}
{\color{green}{  - \sum_{Of} V^{IO}_{n_vf} t^{ef}_{MO} }}
\end{align}
\begin{align}\label{eq:W_vvvc}
   W^{ef}_{am_v} =& V^{ef}_{am_v} 
                 + P (ef) \sum _{n_vg} V^{en_v}_{ag}  \tau^{gf}_{m_vn_v}
{\color{green}{  + P (ef) \sum _{Ng} V^{eN}_{ag}  \tau^{gf}_{m_vN} }}
                 + \sum_g W^{ef}_{ag} t^g_{m_v} \nonumber \\&
                 + \sum_{n_v} \overline{F}^{n_v}_a t^{ef}_{m_vn_v}  
{\color{green}{  + \sum_N \overline{F}^N_a t^{ef}_{m_vN} }}
                 + \sum_{n_v>o_v} V_{am_v}^{n_vo_v} \tau^{ef}_{n_vo_v} 
{\color{green}{  + \sum_{n_vO} V_{am_v}^{n_vO} \tau^{ef}_{n_vO} }}
{\color{green}{  + \sum_{N>O} V_{am_v}^{NO} \tau^{ef}_{NO} }}\nonumber \\&
                 - P(ef) \sum_{n_v} \overline{W}^{n_vf}_{am_v}  t^e_{n_v}
{\color{green}{  - P(ef) \sum_{N} \overline{W}^{Nf}_{am_v}  t^e_N }} 
\end{align}
\begin{align}\label{eq:W_vvvv}
   W^{ef}_{ab} = {} &   V^{ef}_{ab} 
               - P(ef) \sum_{m_v} V^{m_vf}_{ab} t^e_{m_v}
{\color{green}{- P(ef) \sum_M V^{Mf}_{ab} t^e_M }}
               + \sum_{m_v>n_v} V^{m_vn_v}_{ab} \tau ^{ef}_{m_vn_v}\nonumber \\&
{\color{green}{+ \sum_{m_vN} V^{m_vN}_{ab} \tau ^{ef}_{m_vN} }}
{\color{green}{+ \sum_{M>N} V^{MN}_{ab} \tau ^{ef}_{MN} }}  
\end{align}
\begin{align}\label{eq:W_bar_ovvo}
   \overline {W}^{{m_v}b}_{e{j_v}} = {} & V^{{m_v}b}_{e{j_v}} 
                           - \sum_{{n_v}f} V^{{m_v}{n_v}}_{ef} t^{bf}_{{n_v}{j_v}} 
{\color{green}{            - \sum_{Nf} V^{{m_v}N}_{ef} t^{bf}_{N{j_v}} }} 
\end{align}
\begin{align}\label{eq:W_bar_cvvo}
   \overline {W}^{{M}b}_{e{j_v}} = {} & V^{{M}b}_{e{j_v}} 
                           - \sum_{{n_v}f} V^{{M}{n_v}}_{ef} t^{bf}_{{n_v}{j_v}} 
{\color{green}{            - \sum_{Nf} V^{{M}N}_{ef} t^{bf}_{N{j_v}} }} 
\end{align}
\begin{align}\label{eq:W_bar_cvvc}
   \overline {W}^{{M}b}_{e{J}} = {} & V^{{M}b}_{e{J}} 
{\color{green}{            - \sum_{{n_v}f} V^{{M}{n_v}}_{ef} t^{bf}_{{n_v}{J}} }} 
{\color{green}{            - \sum_{Nf} V^{{M}N}_{ef} t^{bf}_{N{J}} }} 
\end{align}